\documentclass[3p]{elsarticle}

    \makeatletter
    \def\ps@pprintTitle{%
      \let\@oddhead\@empty
      \let\@evenhead\@empty
      \let\@oddfoot\@empty
      \let\@evenfoot\@oddfoot
    }
    \makeatother
		
\usepackage{amssymb,amsmath}%,mathtools}%,amsfonts} 
\usepackage{subfigure}      %\usepackage{wrapfig}		
\usepackage{float}					%\usepackage{fullpage}
\usepackage{placeins}	

\def\eeq{\relax}
\def\beq#1#2\eeq{\begin{equation}\label{#1}#2\end{equation}}
\def\bal#1#2\eal{\begin{align}\label{#1}#2\end{align}}
\def\bse#1#2\ese{\begin{subequations}\label{#1}#2\end{subequations}}
\def\ba{\begin{aligned}}   \def\ea{\end{aligned}}

\def\tr{\operatorname{tr}} 
\def\rd{\operatorname{d}} 
\def\re{\operatorname{e}} 
\def\ax{\operatorname{ax}} 
\def\bd{\bf} 

\begin{document} %%%%%%%%%%%%%%%%%%%%%%%%%%%%%%%%%%%%%%%%%%%%%%%%%%%%%%%%%%%%%%%%%%%%%%%%%
 
\begin{frontmatter}
\title{Mechanics of elastic networks}
%Elastic moduli of lattice networks, from stiffest to pentamode}  
\author{ Andrew N. Norris}
\address{Mechanical and Aerospace Engineering, Rutgers University, %\\
Piscataway, NJ 08854-8058, USA}

%\subject{mechanical engineering, structural engineering, mechanics}
%\keywords{lattice,pentamode,elasticity}
%\corres{Andrew Norris\\
%\email{norris@rutgers.edu}}

\begin{abstract} %%%%%%%%%%%%%%%%
We consider a periodic lattice structure in $d=2$ or $3$ dimensions  with  
unit cell  comprising $Z$  thin elastic members emanating from a similarly situated central node.  A general theoretical approach provides an algebraic formula for the effective elasticity of such frameworks. The method  yields the effective cubic elastic constants for 3D space-filling lattices with $Z=4$, $6$, $8$, $12$ and $14$, the latter being the "stiffest" lattice proposed by \cite{Gurtner14}. The analytical expressions provide explicit formulas for the effective properties of pentamode materials, both isotropic and anisotropic, obtained from the general formulation in  the stretch dominated limit for $Z= d+1$.   
\end{abstract}   %%%%%%%%%%%%%%%%
\end{frontmatter}

%\begin{fmtext}
\section{Introduction} \label{sec1}

\begin{figure}[h] %%%%%%%%%%%%%%%%%%%%%%%%%%%%%%%%%%%%%%%%%%%%%%%%%%%%%%%%%%
   \begin{center} 
	\subfigure[$Z=4$]{%    
\includegraphics[width=1.4in,angle=0]{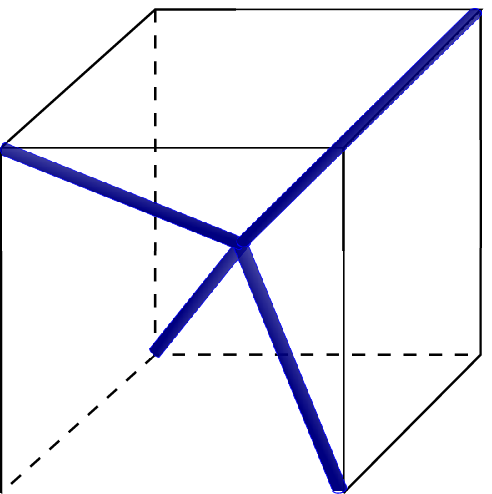}
}
\hspace{0.35in}
	\subfigure[$Z=6$]{%
\includegraphics[width=1.5in,angle=0]{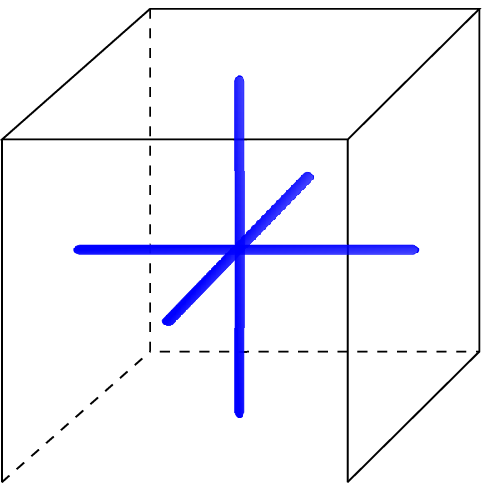}
}%\hspace{-0.32in}
\\
	\subfigure[$Z=8$]{%
\includegraphics[width=1.4in,angle=0]{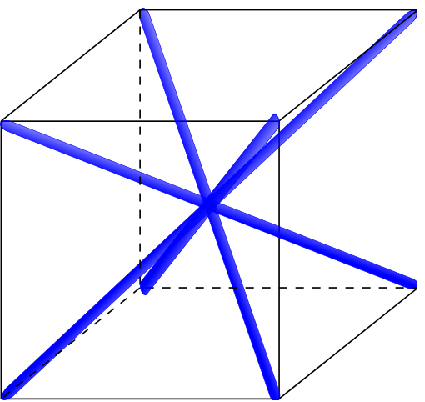}
} \hspace{0.35in}
\subfigure[$Z=12$]{%
\includegraphics[width=1.5in,angle=0]{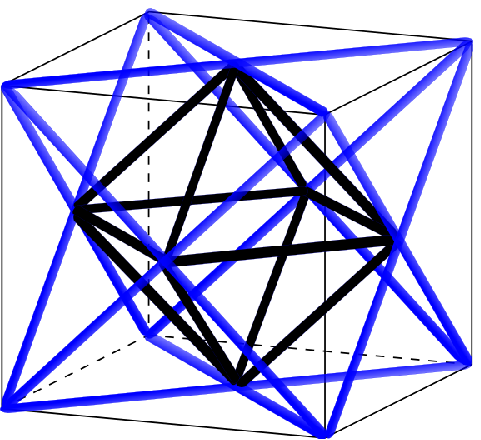}
}
\\
	\subfigure[$Z=14$]{%
\includegraphics[width=1.5in,angle=0]{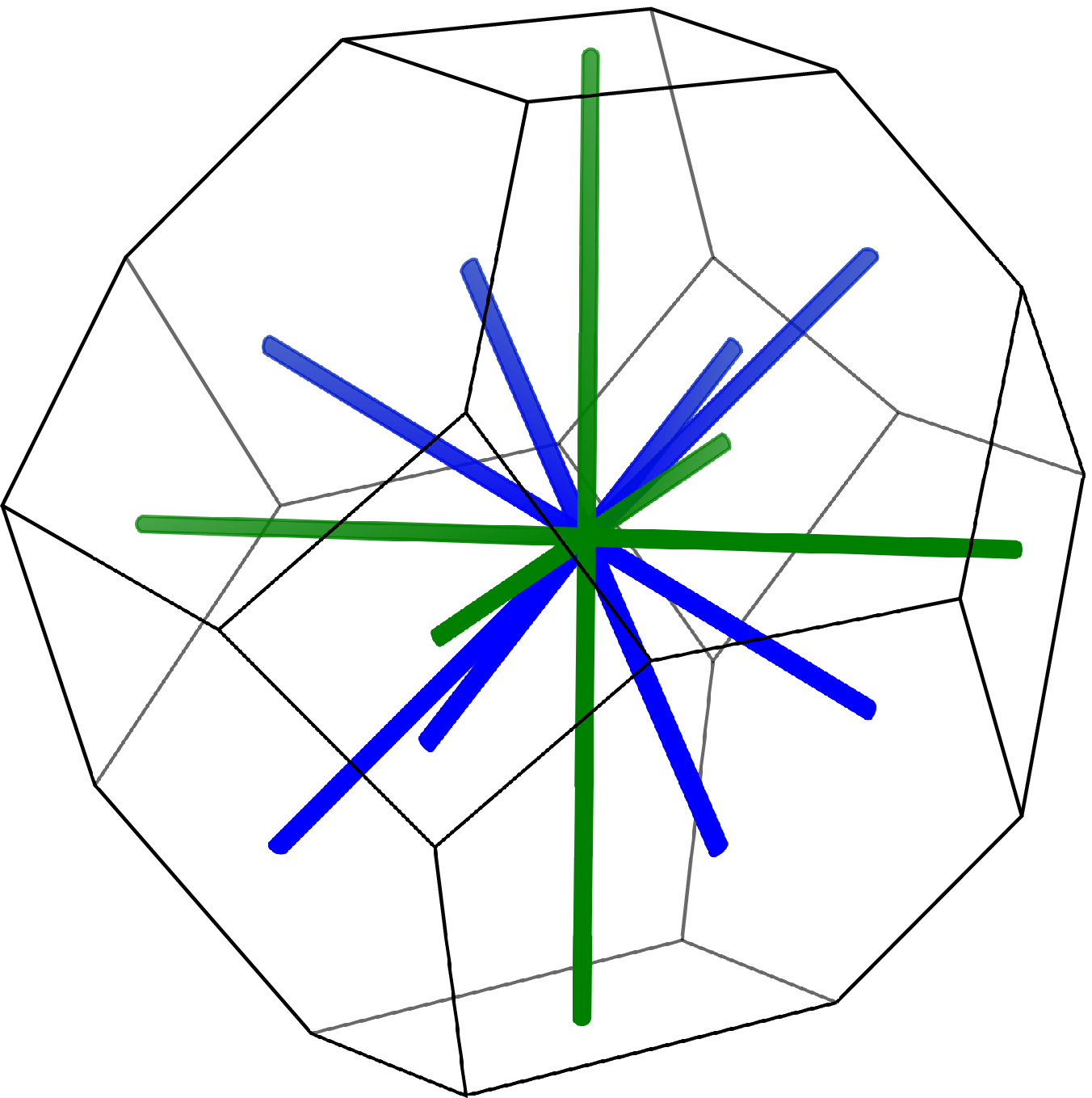}
}
\caption{Unit cell for some  lattices considered, see Table \ref{table}. 
The stretch dominated $Z=14$ lattice with node at the center of the tetrakaidecahedral unit cell has maximal stiffness \cite{Gurtner14}.  Colours are used to illustrate the structure for $Z=12$ and to differentiate the shorter (blue) members from the longer ones(green) for $Z=14$.}
   \label{fig1}
	\end{center}
   \end{figure}

Space frames, or periodic lattice structures of rods and joints, have long been of interest to engineers, architects, materials scientists and others. The octet truss, for instance, which is common in modern large scale structures because of its load bearing capacity  may be attributed to  Alexander Graham Bell's interest in  tetrahedral cells for building man-carrying kites \cite{Bell03}.  Recent fabrication of  micro-architectured materials have  used the octet truss tetrahedral cell design to achieve  ultralight and ultrastiff structures \cite{Fang14}.    An even stiffer structure comprising tetrakaidecahedral unit cells was proposed by  \cite{Gurtner09,Gurtner14}, see Fig.\ \ref{fig1}.  Unlike the octet truss which has cubic elastic symmetry, the tetrakaidecahedral structure can display isotropic effective elastic properties.  At the other end of the stiffness spectrum  for elastic lattice structures are pentamode materials (PMs) with five easy modes of deformation \cite{Milton95}  (see also page 666 of \cite{Milton01}).   The range of such material properties, including high stiffness, strength and fracture toughness, exhibited by  low density micro-architectured 	materials is reviewed in \cite{Fleck10}. 

%\end{fmtext}\maketitle   %%%%%%%%%%%%%%% End of first page %%%%%%%%%%%%%%%%%%%%%
The response of low density lattice structures depends on whether the deformation under load is dominated by stretching versus bending. 
 This in turn depends upon the  coordination number, $Z$, the number of nearest neighboring joints in the unit cell, see  Fig.\ \ref{fig1} for several examples ranging from pentamodal $(Z=4)$ to stiffest $(Z=14)$.   
Maxwell \cite{Maxwell1864}  described  the necessary although not sufficient condition for a $d$-dimensional $(d=2,3)$ space frame of $b$ struts and $j$  pin joints to be just rigid:  $b-3=(j-3)d$.   For an infinite periodic structure,  $b\approx jZ/2$, Maxwell's condition becomes $Z=2d$.
Structures with  $Z=2d$, known as isostatic  lattices,  are at the threshold of mechanical  stability \cite{Lubensky12}.    A closer examination of the issue taking into account the degrees of freedom in the applied strain field, $d(d+1)/2$,  leads to the conclusion that the 
necessary and sufficient condition for
rigidity of frameworks with coordination number $Z$ is  $Z \ge d(d+1)$ \cite{Deshpande01}. The octet truss lattice $(Z=12)$ is an example of a 3D lattice which  satisfies the  rigidity condition  \cite{Deshpande01a}.  Three dimensional  frameworks with  $Z <12$ admit soft modes; thus, as we will see in \S\ref{sec4}\ref{4.6}, a cubic framework with $Z = 6$ has 3 soft modes.    
Zero frequency modes, "floppy" modes, that occur for $Z<2d$ correspond to collapse mechanisms, a topic also examined by  \cite{Hutchinson06} for truss-like 2D lattices. 

Three dimensional   elasticity    is characterized by 6 positive eigenvalues \cite{kelvin}.    A pentamode material (PM) in 3D is the special case of elasticity with five zero eigenvalues, hence "penta".
 An inviscid compressible fluid like water serves as a useful reference material for PMs since it has a single bulk modulus but zero shear rigidity,  the elastic stiffness tensor is  ${\bf C} = K_0\, {\bf I}\otimes {\bf I} $ 
$\Leftrightarrow $ 
$C_{ijkl} = K_0 \,{\delta}_{ij} {\delta}_{kl}$, where $K_0$ is the bulk modulus and $C_{ijkl}$ are the components of ${\bf C}$.  This form of the elastic moduli corresponds to a rank one $6\times 6$ matrix Voigt matrix $[C_{IJ}]$  with  
 single non-zero eigenvalue  $3K_0$. %  since ${\bf C} {\bf I}= 3K_0  {\bf I}$. 
 PMs can therefore be thought of as elastic generalizations of water but without the ability to flow; however, unlike water, for which the stress is isotropic,  PMs can display anisotropy. 
Recent interest in PMs has increased after the observation  that they provide the potential for realizing transformation acoustics \cite{Norris08b}.  
  Pentamode materials can be realized from specific microstructures with tetrahedral-like unit cells \cite{Milton95,Kadic13}.   These types of PM lattice structures are related to low density materials such as foams in which the low density is a consequence of the low filling fraction of the solid phase, see 
\cite{Christensen00} for a  review of mechanical properties of low density materials. 
Here we consider specific microstructures and find explicit values for the elastic moduli for isotropic and anisotropic PMs. 

The purpose of the present paper is two-fold.  First, we fill the need for  a general theoretical approach that provides a simple means to estimate the effective elasticity of frameworks with nodes which are all similarly situated.  Nodes are similarly situated if the framework appears the same when  viewed from any one of the nodes \cite{Nye85}; the unit cell must therefore be space-filling, as are the cases in Fig.\ \ref{fig1}.  Specific homogenization methods  have been proposed for lattice structures;  e.g. \cite{Tollenaere98} use a mix of analytical and finite element methods, while \cite{Martinsson03a,Martinsson03b}   provide a general mathematical scheme that is not easy to implement in practice.   More general micropolar elasticity theories have also been considered for two-dimensional frameworks, e.g. by applying force and moment balances on the unit cell \cite{Wang1999}, or alternatively, using energy based methods \cite{Kumar2004}.
The method proposed  here derives the elastic tensor relating the symmetric stress to the strain.  It does not assume micropolar theories, although the solution  involves a local rotation within the unit cell required for balancing the moments, see \S\ref{sec3}. In contrast to prior works, the present method is explicit and practical; it provides for instance, the  effective cubic elastic constants for all the examples in   Fig.\ \ref{fig1}, see \S\ref{sec4}\ref{4.6}.   The second objective is to provide  analytical expressions for the effective properties of pentamode materials, both isotropic and anisotropic.  The general theory derived here is perfectly suited to this goal.  We show in \S\ref{sec4}\ref{4.1} 
that the minimal coordination number necessary for a fully positive definite elasticity tensor 
is $Z= d+1$ ($d=2$ or $3$), the pentamode limit therefore follows by taking the stretch dominated limit for $Z=d+1$. 

%%%%%%%%%%%%%%

%%%%%%%%%%%%%

The paper  proceeds as follows:  The lattice model is introduced and the main results for the effective properties are summarized in \S\ref{sec2}.  The detailed derivation is presented in \S\ref{sec3}.  In \S\ref{sec4}  some  properties of the effective moduli are described, including the stretch dominated limit, and examples of 5 different lattice structures are given.  Pentamode materials, which  arise as a special case of the stretch dominated limit when the coordination number is $d+1$, are discussed in  \S\ref{sec5}.  
The two dimensional case is presented in \S\ref{sec6} and 
conclusions are given in  \S\ref{sec7}.

\section{Lattice model} \label{sec2}
% \subsection{Setup}
The structural unit cell  in $d$-dimensions ($d=2,3$) comprises $Z\ge d+1$ rods and has  volume $V$.   
 Let ${\bf 0}$ denote the position of  the single junction in the  unit cell with the cell edges at the midpoints of the rods, located at  ${\bf R}_i$  for $i=1,\ldots, Z$.
Under the action of a static loading the relative position of the vertex   initially located at ${\bf R}_i$   moves to ${\bf r}_i$. 
The angle between members $i$ and $j$ before and after deformation is  $\Psi_{ij} =  \cos^{-1} \big({\bf R}_i\cdot {\bf R}_j{/(R_iR_j)} \big)$ and   $ \psi_{ij} =  \cos^{-1} \big({\bf r}_i\cdot {\bf r}_j{/(r_ir_j)} \big)$, respectively, where $R_i=|{\bf R}_i|$, $r_i=|{\bf r}_i|$.  
   The end displacement $\Delta  {\bf r}_i =   {\bf r}_i - {\bf R}_i$ is decomposed as $\Delta  {\bf r}_i = \Delta  {\bf r}_i^\parallel + \Delta  {\bf r}_i^\perp$.  In the linear approximation assumed here  $\Delta  {\bf r}_i^\parallel \approx  \Delta r_i {\bf e}_i$ 	
	where $\Delta r_i = r_i - R_i$ and the unit axial vector is ${\bf e}_i = {\bf R}_i /R_i$  ($|{\bf e}_i|=1$).   The  change  in angle between members $i$ and $j$ is  $\Delta \psi_{ij} \equiv \psi_{ij}- \Psi_{ij}$, $j\ne i$.  The transverse displacement  
	$ \Delta  {\bf r}_i^\perp$ can include a contribution $ \Delta  {\bf r}_i^{\text{rot}}$ $({\bf e}_i\cdot   \Delta  {\bf r}_i^{\text{rot}} =0)$ caused by rigid body rotation of the unit cell. We therefore define
	$\Delta  {\bf r}_i^b =  \Delta  {\bf r}_i^\perp - \Delta  {\bf r}_i^{\text{rot}}$, the transverse displacement associated with flexural bending.   Vectors perpendicular to ${\bf e}_i$ are used to define transverse bending forces: 	
	The unit vector ${\bf e}_{ij}$ is perpendicular to  ${\bf e}_i$ and lies in the plane spanned by  ${\bf e}_i$ and ${\bf e}_j$ with ${\bf e}_{ij}\cdot {\bf e}_j <0$, 
that is\footnote{If   ${\bf e}_i = - {\bf e}_j$ we consider a slight perturbation so that $\psi_{ij}\ne \pi$.} 
%\beq{=34}
${\bf e}_{ij} 
= {\bf e}_i \times 
\big(  {\bf e}_i \times {\bf e}_j \big)\,  |{\bf e}_i \times {\bf e}_j |^{-1 }
=  (\cos \psi_{ij} {\bf e}_i - {\bf e}_j )/ \sin \psi_{ij}
$, $i\ne j \in \overline{1Z}$.
%=  { \csc \psi_{ij}} ( \cos \psi_{ij} {\bf e}_i - {\bf e}_j ) .
%\eeq 
	The unit vector(s) ${\bf e}_i^\alpha$, $\alpha =1:d-1$, are such that $\{  {\bf e}_i, \, {\bf e}_i^\alpha \}$ form an orthonormal set of $d$-vectors.  Summation on repeated lower case Greek superscripts is understood (and only relevant for 3D).  Define
	\beq{-50}
	{\bf P}_i^\parallel = {\bf e}_i \otimes  {\bf e}_i,  \ \
		{\bf P}_i^\perp=  {\bf e}_i^\alpha \otimes {\bf e}_i^\alpha
	 , 
	\eeq
	so that ${\bf P}_i^\parallel + 		{\bf P}_i^\perp = {\bf I}$,  the unit matrix in $d$-dimensions. 
	The axial tensor of a vector ${\bf v}$ is defined by its action on a vector ${\bf w}$ as  $\ax ({\bf v}){\bf w}= {\bf v} \times {\bf w}$. 
	Finally, although the derivation will be mostly coordinate free, for the purpose of defining examples and the components of the effective stiffness tensor,   
	we will use the orthonormal basis $\{ {\bf a}_q\}$ $(q=1:d)$.  

%\clearpage
	      %%%%%%%%%%%%%%%%%%%%%%%%%%%%%%%%%%%%%%%%%%%%%%%%%%%%%%%%%%

 \subsection{Forces on individual members}

The members interact in the static  limit via combined axial forces directed along the members, and bending moments, associated  with axial deformation and transverse flexure, respectively.  
We also include the possibility of nodal bending stiffness, associated with torsional spring effects at the junction The strain energy can then be represented  \cite{Gurtner14}
\beq{--2}
{\cal H} = {\cal H}^s +  {\cal H}^b + {\cal H}^n
\eeq
for stretch, bending and nodal deformation, respectively. 
Later, we  consider the limit in which the contributions from bending,  ${\cal H}^b$ and ${\cal H}^n$, are small, and the deformation may be approximated by axial forces only, which is the stretch-dominated limit.  Physically, this corresponds to slender members with small thickness to length ratio. 
	
	We assume strain energy of the form
\beq{-67}
{\cal H}^s = \sum_{i=1}^Z \frac{  (\Delta r_i)^2}{2M_i}, 
\quad
{\cal H}^b = \sum_{i=1}^Z \frac{(\Delta r_i^b)^2}{2N_i}, 
\quad
{\cal H}^n = \sum_{i=1}^Z\sum_{j\ne i} \frac{R_iR_j} {2N_{ij}}  \, (\Delta \psi_{ij})^2 
\eeq	
	where  $M_i$ are  the   axial compliances, $N_i$  the bending compliances and  
 $N_{ij}$ are the nodal bending  compliances.   
	The force acting at the end of member $i$ ($i=1,\ldots, Z$) is 
\beq{-45}
{\bf f}_i = {\bf f}_i^s +{\bf f}_i^b+{\bf f}_i^n
\eeq
where ${\bf f}_i^s  =\Delta  {\bf r}_i^\parallel  /M_i$, acting parallel to the member, is associated with stretching. 
The perpendicular component of the force   acting on the member's end is comprised of a shear force ${\bf f}_i^b  =  \Delta  {\bf r}_i^b /{N_i}$ caused by the bending of the member, plus a shear force ${\bf f}_i^n =R_j  \Delta \psi_{ij} {\bf e}_{ij} / N_{ij}$ associated with the node compliance.  
The axial and bending compliances can be related to the member properties via   
\beq{=12}
M_i =\int_0^{R_i} \frac{\rd x}{E_iA_i}, 
\quad
N_i= \int_0^{R_i}\frac{x^2\rd x}{E_iI_i} 
, \quad i \in \overline{1Z}
\eeq 
where  $E_i(x)$, $A_i(x)$, $I_i(x)$ are the Young's modulus,  cross-sectional area and moment of inertia, with $x=0$ at the nodal junction.  We assume  circular or square cross-section in 3D so that only a single bending compliance is required for each member, otherwise the results below involving $N_i$ are not generally valid although they could be amended with necessary analytical complication.  The nodal bending  compliances $N_{ij}\ge 0$ are arbitrary and satisfy the  symmetry $N_{ij}=N_{ji}$ which ensures  that the sum of the moments of the node bending  forces are zero.

\FloatBarrier 

\subsection{Effective stress and moduli}

We consider the forces on the members of the unit cell responding to an applied macroscopic loading.   The  forces acting at the node of the unit cell are equilibrated, as are the moments,  
\beq{5-4}
\sum_{i=1}^{Z}  {\bf f}_i =0 , \ \ \
\sum_{i=1}^{Z}  {\bf R}_i \times {\bf f}_i =0 .
\eeq 
Treating the volume of the cell as a continuum with equilibrated stress $\boldsymbol \sigma$, integrating div${{\bf x}\otimes\boldsymbol \sigma } = \boldsymbol \sigma$ over $V$ and identifying the tractions as the point forces $ {\bf f}_i$ acting on the cell boundary, implies the well-known connection 
\beq{2}
{\boldsymbol \sigma} = V^{-1}\sum_{i=1}^{Z}  {\bf R}_i\otimes {\bf f}_i . 
\eeq   
The symmetry of the stress, ${\boldsymbol \sigma}={\boldsymbol \sigma}^T$, is guaranteed by the moment balance \eqref{5-4}$_2$. 
Our aim is to derive the effective elastic moduli defined by the fourth order tensor
${\bf C}$ which relates the stress to the macroscopic strain ${\boldsymbol \epsilon}$ according to 
\beq{-39}
 {\boldsymbol \sigma} = {\bf C}
{\boldsymbol \epsilon} .
\eeq
The elements of the elastic stiffness $\bf C$ when expressed in an orthonormal basis  possess the  symmetries $C_{ijkl} =C_{jikl}$ and $C_{ijkl} =C_{klij}$, and the  elements can also be represented in terms of the Voigt notation via $C_{ijkl}\to C_{IJ}= C_{JI}$.   

\subsection{Summary of the main result for the effective elastic stiffness} \label{sec2.3}

We first introduce the vectors ${\bf d}_i$, ${\bf d}_i^\alpha$, ${\bf d}_{ij}$ $( = {\bf d}_{ji} 
) $, the 
second order symmetric tensors ${\bf D}_i$,  ${\bf D}_i^\alpha$, ${\bf D}_{ij}$ $( = {\bf D}_{ji} )  $ 
and the $L\times L$ matrix with elements $P_{ij}$, where  $L=Z d +Z(Z-1)/2$:
\bse{11}
\bal{7=1}
{\bf d}_i &= \frac{ {\bf e}_i}{\sqrt{M_i}} ,  
\qquad 
{\bf d}_i^\alpha = \frac{ {\bf e}_i^\alpha}{\sqrt{N_i}}, \ \  (\alpha =1:d-1)
\qquad 
{\bf d}_{ij} = \sqrt{ \frac{R_iR_j}{N_{ij}}} \,  
\Big( \frac{{\bf e}_{ij}}{R_i} +\frac{{\bf e}_{ji}}{R_j} \Big) ,   
\\ 
{\bf D}_i  &=  
\frac{ R_i {\bf P}_i^\parallel }{\sqrt{VM_i }}  ,  
\ \ 
{\bf D}_i^\alpha =   
\frac{ R_i }{\sqrt{VN_i }} \frac 12
( {\bf e}_i \otimes {\bf e}_i^\alpha + {\bf e}_i^\alpha \otimes {\bf e}_i),  
\ \
{\bf D}_{ij}   = 
 \sqrt{ \frac{R_iR_j}{VN_{ij}}} \,  
\big( 
{\bf e}_i\otimes  {\bf e}_{ij}
+ {\bf e}_j\otimes {\bf e}_{ji} \big) 
\label{7=4}
\\ 
\label{7=6}
\left. \{ {\bf u}_k \}\right|_{k=1}^L &= \{  {\bf d}_i ,\,  {\bf d}_i^\alpha, \, {\bf d}_{ij}  \}, 
\ \ \ 
\left. \{ {\bf U}_k \}\right|_{k=1}^L = \{  {\bf D}_i ,\, {\bf D}_i^\alpha ,  \, {\bf D}_{ij}  \} , 
\ \  (\alpha =1:d-1) 
\\
\label{7=8} 
P_{ij} &= \delta_{ij} - {\bf u}_i\cdot \Big(
\sum_{k=1}^{L}  {\bf u}_k\otimes {\bf u}_k  \Big)^{-1} \cdot {\bf u}_j,\ \ i,\, j = 1: L,
\eal
\ese
then, under  some general assumptions applicable to the 3D structures in Fig.\ \ref{fig1}, eq.\ \eqref{3=0}, the  effective moduli can be be written  
\beq{7=9}
{\bf C} =   \sum_{i, j=1}^{L}
 P_{ij}  {\bf U}_i\otimes {\bf U}_j  .
\eeq
These results are derived in the next section and  implications  are discussed in \S\ref{sec4}, including a simple expression \eqref{9=7} for the elastic moduli represented in $6\times 6$ Voigt notation.    The general structure  of eqs. \eqref{11} holds for $d=2$ without  requiring the zero rotation conditions of eq.\ \eqref{3=0}, as discussed in \S\ref{sec6}.

\section{Derivation of the effective elasticity tensor} \label{sec3}

\subsection{Affine deformation}

Strain is introduced through the  affine kinematic assumption that the effect of  deformation is to cause the cell edges to displace  in a linear manner proportional to the (local) deformation gradient ${\bf F}$. 
Edge points originally located at ${\bf R}_i$ are translated to ${\bf F}{\bf R}_i$.  In addition to the affine motion, we include two $d$-vectors, introduced to satisfy the equilibrium conditions  \eqref{5-4}. Following \cite{Wang00} we assume that  the junction moves from the origin to  ${\boldsymbol \chi}$.  An additional rotation ${\bf Q} \in $SO$(d)$ is introduced, so that the vector defining the edge relative to the vertex is  
\beq{3}
{\bf r}_i= {\bf Q}{\bf F}{\bf R}_i - {\boldsymbol \chi}.
\eeq
 
  The linear approximation for the deformation is 
${\bf F} = {\bf I} +{\boldsymbol \epsilon} + {\boldsymbol \omega}$ with ${\boldsymbol \epsilon} = {\boldsymbol \epsilon}^T$ and 
${\boldsymbol \omega} = - {\boldsymbol \omega}^T$.   We take ${\bf Q} = \re^{\boldsymbol \Gamma}
= {\bf I} + {\boldsymbol \Gamma} + $O$({\boldsymbol \Gamma})^2$ where the skew symmetric matrix 
${\boldsymbol \Gamma} $ is defined by the $d$-vector ${\boldsymbol \gamma}$ as
${\boldsymbol \Gamma} = \ax ({\boldsymbol \gamma})$. 
 Hence, 
\beq{-5}
\Delta {\bf r}_i \equiv  {\bf r}_i - {\bf R}_i  = ({\boldsymbol \epsilon} + {\boldsymbol \omega}+{\boldsymbol \Gamma}) {\bf R}_i - {\boldsymbol \chi} . 
\eeq
In the linear approximation ${\bf r}_i$ can equally well be taken along ${\bf R}_i$ as far as second order terms are concerned.  
Thus, 
\beq{-51}
\ba
\Delta {\bf r}_i^\parallel   &= \big( R_i  {\bf P}_i^\parallel : {\boldsymbol \epsilon}  - {\bf e}_i\cdot {\boldsymbol \chi} \big)\, {\bf e}_i, 
\\
\Delta {\bf r}_i^\perp &= 
{\bf P}_i^\perp
( R_i   {\boldsymbol \epsilon}{\bf e}_i - {\boldsymbol \chi}  ) 
+ R_i ({\boldsymbol \omega} +{\boldsymbol \Gamma}) {\bf e}_i ,
%- (  {\bf I}- {\bf e}_i\otimes {\bf e}_i ) {\boldsymbol \chi} 
\\
\Delta  \psi_{ij}  
&= {\bf e}_i\cdot  {\boldsymbol \epsilon} {\bf e}_{ij}
+ {\bf e}_j\cdot  {\boldsymbol \epsilon} {\bf e}_{ji}
-  \big( R_i^{-1}{{\bf e}_{ij}} +R_j^{-1}{{\bf e}_{ji}} \big)\cdot {\boldsymbol \chi} .
\ea
\eeq
The tangential displacement governing the shear bending force is, after removing the affine rigid body rotation, 
\bal{-59}
\Delta  {\bf r}_i^b  = 
\Delta {\bf r}_i^\perp - R_i {\boldsymbol \omega} {\bf e}_i 
. 
\eal
Note that we retain the unknown rotation ${\boldsymbol \Gamma}$ in order to satisfy the moment equilibrium condition \eqref{5-4}$_2$. 
Hence,  in the linear approximation \eqref{-45} becomes
\bal{43}
{\bf f}_i  &= M_i^{-1 } {R_i}  ({\bf P}_i^\parallel : {\boldsymbol \epsilon})\,  {\bf e}_i 
+ N_i^{-1 } {R_i} \big( {\bf P}_i^\perp{\boldsymbol \epsilon}{\bf e}_i +  {\boldsymbol \gamma}\times {\bf e}_i \big) 
+ \sum_{j\ne i} N_{ij}^{-1 }{ R_j}   \big( 
{\bf e}_i\cdot  {\boldsymbol \epsilon} {\bf e}_{ij}
+ {\bf e}_j\cdot  {\boldsymbol \epsilon} {\bf e}_{ji} \big) {\bf e}_{ij} 
\nonumber \\ 
& \quad 
-
\Big(   M_i^{-1 }{\bf P}_i^\parallel
+ N_i^{-1 } {\bf P}_i^\perp 
+ \sum_{j\ne i} N_{ij}^{-1} R_j  {\bf e}_{ij} \otimes 
\big( R_i^{-1}{{\bf e}_{ij}} +R_j^{-1}{{\bf e}_{ji}} \big)
\Big)
 {\boldsymbol \chi} . 
\eal
This explicit expression for the forces allows us to determine the vectors ${\boldsymbol \chi}$ and ${\boldsymbol \gamma}$, next. 

\subsection{Solution of the equilibrium equations}

Consider first the moment balance condition \eqref{5-4}$_2$. Of the three terms comprising the force in eq.\ \eqref{-45} only the bending shear forces ${\bf f}_i^b$ does not automatically yield zero moment.  Equilibrium of the moments therefore reduces to  % \end{document}
\beq{4=6}
\sum_{i=1}^{Z}  {\bf R}_i \times {\bf f}_i^b =0 .
%\ \ \Leftrightarrow  \ \ 
\eeq
Substituting ${\bf f}_i^b  =  \Delta  {\bf r}_i^b /{N_i}$ and using  eqs.\ \eqref{-59} and  \eqref{4=6} allows us to find 
${\boldsymbol \gamma}$ in the form 
\beq{4=7}
{\boldsymbol \gamma}= {\bf B} 
\Big( {\bf g} \times  {\boldsymbol \chi}  -
\sum_{j=1}^{Z}  \frac{R_j^2}{N_j} {\bf e}_j \times   {\boldsymbol \epsilon}{\bf e}_j \Big)
\ \text{where} \ 
{\bf B} = \Big(
\sum_{i=1}^{Z}  \frac{R_i^2}{N_i} {\bf P}_i^\perp \Big)^{-1} , 
\ \ 
 {\bf g} = \sum_{i=1}^{Z}  \frac{R_i}{N_i} {\bf e}_i . 
\eeq
The force on member $i$ becomes, using eq.\ \eqref{43}, 
\bal{431}
{\bf f}_i  &= \frac{R_i}{M_i}  ({\bf P}_i^\parallel : {\boldsymbol \epsilon})\,  {\bf e}_i 
+ \frac{R_i}{N_i} \big( {\bf P}_i^\perp {\boldsymbol \epsilon}{\bf e}_i +  
 \ax({\bf e}_i ) {\bf B}
\sum_{j=1}^{Z}  \frac{R_j^2}{N_j} \ax({\bf e}_j )   
 {\boldsymbol \epsilon}{\bf e}_j  \big)
+ \sum_{j\ne i} \frac{R_j}{N_{ij}}   \big( 
{\bf e}_i\cdot  {\boldsymbol \epsilon} {\bf e}_{ij}
+ {\bf e}_j\cdot  {\boldsymbol \epsilon} {\bf e}_{ji} \big) {\bf e}_{ij} 
\nonumber \\ 
& \quad 
-
\Big(   \frac{{\bf P}_i^\parallel }{M_i}
+\frac{ {\bf P}_i^\perp }{N_i}
+    \frac{R_i}{N_i} \ax({\bf e}_i ) {\bf B}
 \ax({\bf g} )
+ \sum_{j\ne i} \frac{R_j}{N_{ij}}   {\bf e}_{ij} \otimes 
\big( \frac{{\bf e}_{ij}}{R_i}  +\frac{{\bf e}_{ji}}{R_j} \big)
\Big)
 {\boldsymbol \chi} . 
\eal

The equilibrium condition \eqref{5-4}$_1$ can then be solved for  ${\boldsymbol \chi}$  as  
\beq{-10}
\ba
 {\boldsymbol \chi}
 =  {\bf A}^{-1} \sum_{i=1}^{Z}   
 \Big( &  \frac {R_i}{M_i }  ({\bf P}_i^\parallel : {\boldsymbol \epsilon})\, {\bf e}_i 
+  \frac{R_i}{N_i} \big( {\bf P}_i^\perp +  
 \ax({\bf g} ) {\bf B}
 R_i \ax({\bf e}_i )      \big) \cdot {\boldsymbol \epsilon}{\bf e}_i
\\ &
+ \sum_{j\ne i} \frac{ R_j}{N_{ij} }   \big( 
{\bf e}_i\cdot  {\boldsymbol \epsilon} {\bf e}_{ij}
+ {\bf e}_j\cdot  {\boldsymbol \epsilon} {\bf e}_{ji} \big) {\bf e}_{ij}   \Big)
\ea
\eeq
{where} 
\beq{304}
{\bf A} = 
\sum_{i=1}^{Z}   \Big(  \frac{{\bf P}_i^\parallel }{M_i}
+\frac{ {\bf P}_i^\perp }{N_i}
+ \sum_{j\ne i} \frac {R_iR_j}{N_{ij} } \frac{{\bf e}_{ij}}{R_i} \otimes \big( 
 \frac{{\bf e}_{ij}}{R_i}
+\frac{{\bf e}_{ji}}{R_j} \big) \Big)
+     \ax({\bf g} ) {\bf B}  \ax({\bf g} )  .
\eeq
Equations \eqref{2}, \eqref{431} and \eqref{-10}  provide the desired linear relation between the strain and the stress from which one can derive the effective elastic moduli. 

\subsection{A simplification}
While eqs.\ \eqref{2}, \eqref{-39},  \eqref{431}-\eqref{304}  provide all of the necessary ingredients for the most general situation we assume for the remainder of the paper that the unit cell rotation vanishes, implying  ${\boldsymbol \gamma} = 0$.  Hence, the vector ${\bf g}$ and the second term in   the  
expression for  ${\boldsymbol \gamma} $ in eq.\ \eqref{4=7} vanish. 
  The latter identity  is equivalent to 
$
({\bf D}{\bf v} )\times {\bf v}=0 $ $\forall {\bf v}$
where $ {\bf D}=
 \sum_{i=1}^{Z}  {R_i^2}N_i^{-1} \, {\bf e}_i \otimes {\bf e}_i$.
This implies that ${\bf D}$ must be proportional to the identity, hence   the zero rotation condition
 may be written  
\beq{3=0}
 \sum_{i=1}^{Z}    \frac{R_i}{N_i} {\bf e}_i  = 0
\ \ \text{and} \ \  
\sum_{i=1}^{Z}  \frac{R_i^2}{N_i} \big( {\bf e}_i \otimes {\bf e}_i - 
\frac 1d  {\bf I}\big) = 0  
\ \ \Leftrightarrow  \ \text{zero cell rotation}. 
\eeq
The identities \eqref{3=0} hold  for  the examples considered later.  Note that the assumption of zero rotation is not necessary for stretch dominated lattices in which bending effects are negligible.

\subsection{Effective stiffness}

In order to arrive at an explicit expression for the elastic stiffness tensor we first 
write the stress in terms of strain, using eqs.\  \eqref{2},  \eqref{43}-\eqref{-10}, 
\bal{7=3}
{\boldsymbol \sigma} =& \frac 1V
\sum_{i=1}^{Z}   
\Big( 
\frac{R_i^2}{M_i}  \,{\bf P}_i^\parallel  ({\bf P}_i^\parallel : {\boldsymbol \epsilon})
+ \frac{R_i^2}{N_i}  \, {\bf e}_i \otimes {\bf e}_i^\alpha  ({\bf e}_i^\alpha\cdot  {\boldsymbol \epsilon} {\bf e}_i)
\Big) 
+ \sum^{Z}_{\overset{i=1}{j\ne i}}
\frac{R_iR_j}{N_{ij}}  {\bf e}_i \otimes {\bf e}_{ij} 
\, \big( 
{\bf e}_i\cdot  {\boldsymbol \epsilon} {\bf e}_{ij}
+ {\bf e}_j\cdot  {\boldsymbol \epsilon} {\bf e}_{ji} \big)
\notag \\
& - \frac 1V\Big( 
\sum_{i=1}^{Z}   \big(
\frac{R_i}{\sqrt{M_i }} \,{\bf P}_i^\parallel\, {\bf d}_i 
+\frac{R_i}{\sqrt{N_i }} \, ({\bf e}_i \otimes {\bf e}_i^\alpha )\, {\bf d}_i^\alpha
\big)
+\sum^{Z}_{\overset{i=1}{j\ne i}}
 \sqrt{ \frac{R_iR_j}{N_{ij}}} \,
 ({\bf e}_i \otimes {\bf e}_{ij} )\, {\bf d}_{ij} 
\Big) \cdot {\bf A}^{-1} 
\notag \\
& \cdot 
\Big( 
\sum_{k=1}^{Z}  \big(  {\bf d}_k\, 
\frac{R_k}{\sqrt{M_k }} \,{\bf P}_i^\parallel : {\boldsymbol \epsilon}
+ {\bf d}_k^\alpha \, 
\frac{R_k}{\sqrt{N_k }} \, {\bf e}_k^\alpha \cdot  {\boldsymbol \epsilon}{\bf e}_k 
\big)
+\frac 12 \sum^{Z}_{\overset{k=1}{l\ne k}} {\bf d}_{kl} \, 
 \sqrt{ \frac{R_kR_l}{N_{kl}}} 
\big( 
{\bf e}_k\cdot  {\boldsymbol \epsilon} {\bf e}_{kl}
+ {\bf e}_l\cdot  {\boldsymbol \epsilon} {\bf e}_{lk} \big)\Big) .
\eal
It follows from 
\eqref{7=3},  the symmetry of the stress and strain and from the definition of the  second order symmetric tensors
${\bf D}_i$, ${\bf D}_{ij}$ in \eqref{7=4} that  the elastic moduli can be expressed 
\bal{7=5}
 {\bf C} =&
\sum_{i=1}^{Z}   \big( {\bf D}_i \otimes {\bf D}_i + {\bf D}_i^\alpha \otimes {\bf D}_i^\alpha \big)  
+ \frac 12 \sum^{Z}_{\overset{i=1}{j\ne i}}    {\bf D}_{ij} \otimes  {\bf D}_{ij} 
\notag 
\\
& - 
\Big( 
\sum_{i=1}^{Z}    \big( {\bf D}_i \,  {\bf d}_i + {\bf D}_i^\alpha \otimes {\bf d}_i^\alpha \big)
+  \frac 12 \sum^{Z}_{\overset{i=1}{j\ne i}}    {\bf D}_{ij} \,   {\bf d}_{ij} 
\Big) 
\cdot {\bf A}^{-1}  
\notag \\ & 
\cdot 
\Big( 
\sum_{k=1}^{Z}  \big(  {\bf d}_k \,  {\bf D}_k 
+ {\bf d}_k^\alpha \otimes {\bf D}_k^\alpha \big)
+  \frac 12 \sum^{Z}_{\overset{k=1}{l\ne k}}    {\bf d}_{kl} \,   {\bf D}_{kl} 
\Big) .
\eal
Finally, we note, based on the definitions of the vectors in \eqref{7=1}, that 
\beq{7=2}
{\bf A} = 
\sum_{i=1}^{Z}   \big( {\bf d}_i \otimes {\bf d}_i + %\sum_{\alpha=1,2}
{\bf d}_i^\alpha \otimes {\bf d}_i^\alpha \big)
+ \frac 12  \sum^{Z}_{\overset{i=1}{j\ne i}}    {\bf d}_{ij} \otimes  {\bf d}_{ij}  
= \sum^{L}_{i=1}    {\bf u}_i \otimes  {\bf u}_i.
\eeq

The sets $\{ {\bf u}_k \}$ and $\{ {\bf U}_k \}$  defined in \eqref{7=6} combine the $Z$ vectors/tensors associated with stretch, 
the $(d-1)Z$ vectors/tensors associated with shear, and 
  the $Z(Z-1)/2$   vectors/tensors associated with   nodal bending into sets of $L= d Z+ Z(Z-1)/2$ elements in terms of which \eqref{7=5} becomes
\beq{7=7}
 {\bf C} =
\sum_{i=1}^{L}   {\bf U}_i \otimes {\bf U}_i  - 
\big( 
\sum_{i=1}^{L}   {\bf U}_i \,  {\bf u}_i  \big) 
\cdot 
\big( 
\sum_{j=1}^{L}   {\bf u}_j \otimes  {\bf u}_j  \big)^{-1} 
  \cdot 
\big( 
\sum_{k=1}^{L}   {\bf u}_k \,  {\bf U}_k \big) .
\eeq
It then follows from the definition of ${\bf P}$ in \eqref{7=8} that ${\bf C}$ can be expressed   in the form \eqref{7=9}.

\section{Properties of the effective moduli} \label{sec4}

\subsection{Generalized Kelvin form}  \label{4.1}

The $L\times L$ symmetric matrix ${\bf P}$  with elements $P_{ij}$ defined in   eq.\ \eqref{7=8} has the crucial properties 
\beq{7-1}
{\bf P}^2 = {\bf P}, \quad 
\text{rank} \,{\bf P}  = L-d,  
\eeq
i.e.  ${\bf P}$ is a projector, and  the dimension of its projection space is tr\,${\bf P}=L-d$. Hence, the summation in \eqref{7=9} is essentially the sum of $L-d$ tensor products of second order tensors.  This is to be compared with the Kelvin form for the elasticity tensor \cite{kelvin} 
\beq{71-}
{\bf C} = \sum_{i=1}^{3d-3} \lambda_i\, {\bf S}_i \otimes {\bf S}_i
\ \ 
\text{where}\ \  \lambda_i >0, \ \ \tr {\bf S}_i {\bf S}_j = \delta_{ij}. 
\eeq
The second order symmetric tensors are eigenvectors $\{ {\bf S}_i\}$ that diagonalize the elasticity tensor, with eigenvectors $\lambda_i$ known as the Kelvin stiffnesses.   Equation \eqref{7=9} provides a non-diagonal representation for ${\bf C}$.

Note that  $L\equiv L_\text{s}+L_\text{b}+L_\text{n}$ where   
$L_\text{s} = Z$ is associated with stretch, $L_\text{b} = (d-1)Z$ with  bending shear and $L_\text{n} = Z(Z-1)/2 $
with   nodal bending.   A necessary although not sufficient condition for positive definiteness of ${\bf C}$ is that the rank of ${\bf P}$, which is $L-d$, exceed $3d-3$.  
Ignoring nodal bending $(L=L_\text{s}+L_\text{b} )$ this  is satisfied if  $Z \ge d+1 $ for $d=2$ and $3$.  The requirement is stricter in the stretch dominated limit $(L=L_\text{s})$:  $Z \ge 6$ in 2D and $Z \ge 10$ in 3D.

\subsection{$6\times 6$ matrix in 3 dimensions}

The main result of eq.\ \eqref{7=9} implies a simple representation for the  $6\times 6$ matrix   of  elastic moduli %$[C]_{6\times 6}$  
$[C_{IJ}]$ based on the compact Voigt notation $(C_{ijkl}\to C_{IJ})$  in the orthonormal basis $\{{\bf a}_1$, ${\bf a}_2$, ${\bf a}_3\}$.  Let $[u]_{3\times L}$ denote the $L$ vectors $\{ {\bf u}_k\}$ and let 
$[U]_{6\times L}$ denote the $L$ second order tensors $\{ {\bf U}_k\}$ according to 
  $U_{Ik} = {\bf a}_i\cdot {\bf U}_k \cdot {\bf a}_j$     
with the  standard correspondence $ I\in \{1,2,3,4,5,6\} \to ij\in\{11,22,33,23,31,12\}$.
Then eq.\ \eqref{7=9} becomes
\beq{9=7}
\begin{pmatrix}
C_{11} & C_{12} & C_{13} & C_{14} &C_{15} &C_{16} 
\\
C_{12} & C_{22} & C_{23} &C_{24} &C_{25} &C_{26} 
\\
C_{13} & C_{23} & C_{33} &C_{34} &C_{35} &C_{36} 
\\
C_{14} &C_{24} &C_{34}  &C_{44} &C_{45} &C_{46} 
\\
C_{15} &C_{25} &C_{35}  &C_{45} &C_{55} & C_{56} 
\\
C_{16} &C_{26} &C_{36}  &C_{46} &C_{56}&C_{66}  
\end{pmatrix} =  [U][U]^T - [U][u]^T\,    \big([u][u]^T\big)^{-1}\, [u][U]^T .
\eeq

\subsection{Bulk modulus} 

If the effective medium has isotropic or cubic symmetry then a strain 
$
{\boldsymbol \epsilon}
= \varepsilon {\bf I}$ produces strain $ {\boldsymbol \sigma} = d K \varepsilon {\bf I}$ where $K$ is the $d-$dimensional bulk modulus.  More generally, whether or not the symmetry is cubic or isotropic, we can define
$
K = d^{-2} C_{ii jj}$.  The bulk modulus follows from eqs.\ \eqref{11} and \eqref{7=9} as 
\beq{8-}
K=   \frac 1{d^2 V} \sum_{i, j=1}^{Z}
 P_{ij}  \, \frac{R_iR_j}{\sqrt{M_iM_j}} .
\eeq
This simplifies further under the broad assumption that 
\beq{999}
\sum_{i=1}^{Z}\frac{ {\bf e}_i}{\sqrt{M_1}} = 0, 
\eeq
certainly true of all the examples of Fig.\ \ref{fig1} considered in \S\ref{sec4}\ref{4.6}, so that 
\beq{7-}
K=   \frac 1{d^2 V} \sum_{i=1}^{Z}
    \, \frac{R_i^2}{M_i} .
\eeq
Note that the bulk modulus depends only on the axial stiffness of the members. 

Assume the members are the same material $(E_i = E)$, and each has constant cross-section (area or width) $A_i$, then according to eqs.\ \eqref{=12}$_1$ and  \eqref{7-}, 
\beq{=7-1}
K=   \frac \phi {d^2 }  E 
\ \ \text{where} \ 
\phi =   \frac 1{V} \sum_{i=1}^{Z} A_i R_i 
\eeq
is the volume fraction of solid material in the lattice. 
The scaling of bulk modulus with volume fraction, $K \propto \phi E$, is well known, e.g. 
\cite[eq.\ (2.2)]{Christensen95} for $d=2$,  \cite{Warren97},  \cite{Zhu97} for tetrakaidecahedral unit cells (see below), and \cite{Gurtner14}. 

\subsection{Model simplification}
  
While the model considered  is quite general, in practice there is little information on the form of the nodal compliances for practical situations.   For the remainder of the paper we concentrate on just the stretch and shear bending effects, so that $L = L_\text{s}+L_\text{b}+L_\text{n}\to L_\text{s}
+L_\text{b} = dZ$.  The  stress-strain relation is then 
\beq{465}
{\boldsymbol \sigma}  = 
\sum_{i=1}^Z \Big(
 {\bf R}_i \otimes  {\bf X}_i 
\big[ {\bf I}\otimes  {\bf R}_i
-  
\big( \sum_{k=1}^Z    {\bf X}_k \big)^{-1}
\sum_{j=1}^{Z}  
{\bf X}_j 
\otimes {\bf R}_j 
\big] 
\Big) : {\boldsymbol \epsilon} 
\  \text{where} \ 
{\bf X}_i =      \frac{{\bf P}_i^\parallel   }{M_i}
+\frac{ {\bf P}_i^\perp }{N_i}  .
\eeq
 A further simplifications is obtained by ignoring shear bending effects, i.e. 
$L\to L_\text{s} = Z$,  the {\it stretch dominated limit}, considered next. 

\subsection{Stretch dominated limit}\label{3.5}

In this  limit the forces ${\bf f}_i$ have no transverse  components.    Physically, this corresponds to infinite bending compliances, $1/N_{i} = 0$,  $1/N_{ij} = 0$, and may be achieved approximately by long slender members.  
By ignoring shear and nodal bending   
the  expression  for ${\bf C}$ reduces to 
\beq{9=2} 
\ba
{\bf C} = \frac 1V \sum_{i, j=1}^{Z}
 \frac{R_iR_j}{\sqrt{M_iM_j}}\, P_{ij} \, 
{\bf P}_i^\parallel \otimes  {\bf P}_j^\parallel,  \ \ 
%\text{where}  \\
P_{ij} = \delta_{ij} - \frac{{\bf e}_i}{\sqrt{M_i}} \cdot \big(
\sum_{k=1}^{Z}  \frac{{\bf P}_k^\parallel}{M_k}  \big)^{-1} \cdot\frac{{\bf e}_j}{\sqrt{M_j}} .
\ea
\eeq
It follows from eq.\ \eqref{7-1} that the  $Z\times Z$ projection  matrix ${\bf P}$ with elements $P_{ij}$ 
has rank $Z-d$.

%%%%%%%%%%%%%%%%%%%%%%%%%%%%%%%%%%%%%%%%%%%%%%%%%%%%%%%%%%%%%%
\subsection{Examples in 3D: $Z=4, 6, 8, 12, 14$}\label{4.6}

\begin{center}
\begin{table}[h]
\caption{3D lattice structures considered.  They display  cubic elastic symmetry with $C_{11}=K+\frac 43 \mu_2$, $C_{12}=K-\frac 23 \mu_2$ and $C_{44}= \mu_1$ where $K$ is given by eq.\ \eqref{-88}. All cases except  $Z=14$ have uniform rod length $R$ and compliances $M$, $N$.    The boundary of  the tetrakaidecahedral   $(Z=14)$ unit cell  has  36 edges each of length $a$ and  the cell comprises members  of two types: 6 of length $R_1 = {\sqrt{2}} a$ and 8 of length 
 $R_2 = \sqrt{\frac 32}a$, the average length of the  members is $\overline R =1.306\, a$.  The associated  compliances are $M_1$, $N_1$ and $M_2$, $N_2$. 
 The volume fraction $\phi$ in all cases is   based on cylindrical rods of uniform radius $b$.    Note that the volume fraction increases with coordination number $Z$. 
}
\begin{tabular}{c p{1.4cm} c c c c c }   %{||c|p{1.4cm}|c|c|c|c|c ||}
\hline
%&&&&&& \\
Z & cell & V & $\phi$ & $\{ {\bf e}_i\}$ \ (not normalized)  & $\frac{\mu_1}K$ & $\frac{\mu_2}K$
\\  %&&&&&& \\
\hline 
%\hline  &&&&&& \\
4 & diamond& $\frac{64}{3\sqrt{3}}R^3$ & 1.02 $\frac{b^2}{R^2}$ & 
  $\footnotesize \begin{pmatrix} -1\\ -1\\ -1\end{pmatrix}$
  $\footnotesize \begin{pmatrix} -1\\  1\\  1\end{pmatrix}$
  $\footnotesize \begin{pmatrix}  1\\ -1\\  1\end{pmatrix}$
  $\footnotesize \begin{pmatrix}  1\\  1\\ -1\end{pmatrix}$ 
	   & $\frac{9M}{4M+2N}$ & $\frac{3M}{2N}$
\\  %&&&&&& \\ %\hline  &&&&&& \\
6 & simple cubic& $8 R^3$ &  2.36 $\frac{b^2}{R^2}$ & 
$\footnotesize \begin{pmatrix} \pm 1\\ 0\\ 0\end{pmatrix}$
$\footnotesize \begin{pmatrix} 0\\ \pm 1\\ 0\end{pmatrix}$
$\footnotesize \begin{pmatrix} 0\\ 0\\ \pm 1\end{pmatrix}$ &$\frac{3M}{2N}$& $\frac{3}{2}$
\\  %&&&&&& \\ %\hline  &&&&&& \\
8 & BCC& $\frac{32}{3\sqrt{3}}R^3$ &  4.08 $\frac{b^2}{R^2}$ & 
$\footnotesize \begin{pmatrix} \pm 1\\ \pm 1\\ \pm 1\end{pmatrix}$  &$1 +\frac{M}{2N}$ & $\frac{3M}{2N}$
\\  %&&&&&& \\ %\hline  &&&&&& \\
12 & FCC octet truss& $4\sqrt{2} R^3$  & 6.66 $\frac{b^2}{R^2}$ & 
$\footnotesize \begin{pmatrix} 0\\ \pm 1\\ \pm 1\end{pmatrix}$
$\footnotesize \begin{pmatrix} \pm 1\\ 0\\ \pm 1\end{pmatrix}$
$\footnotesize \begin{pmatrix} \pm 1\\ \pm 1\\ 0\end{pmatrix}$  &$\frac 34 + \frac{3M}{4N}$  &$\frac 38 + \frac{9M}{8N}$
\\  %&&&&&& \\ %\hline  &&&&&& \\
%\multirow{2}{*} 
14 & tetrakai- decahedral %\\ tetrakaidecahedral
& $8\sqrt{2} a^3$  & 8.66  $\frac{b^2}{{\overline R}^2}$ & 
$\footnotesize \begin{pmatrix} \pm 1\\ 0\\ 0\end{pmatrix}$
$\footnotesize \begin{pmatrix} 0\\ \pm 1\\ 0\end{pmatrix}$
$\footnotesize \begin{pmatrix} 0\\ 0\\ \pm 1\end{pmatrix}$
$\footnotesize \begin{pmatrix} \pm 1\\ \pm 1\\ \pm 1\end{pmatrix}$  &
$\frac 32 \frac{\frac 1{M_1} + \frac 1{N_2}}{\frac 1{M_1} + \frac 1{M_2}}$
& 
$\frac{\frac 1{N_2} + \frac 2{M_2}+ \frac 3{N_1}}{2\big( \frac 1{M_1} + \frac 1{M_2} \big) }$
%\\  &&&&&& 
\\ \hline %\hline 
\end{tabular}
\label{table}
\end{table}\end{center}

%%%%%%%%%%%%%%%%%%%%%%%%%%%%%%%

All examples display cubic symmetry, with three independent elastic moduli: $C_{11}$, $C_{12}$ and $C_{44}$. 
Introduce the fourth order tensors ${\mathbb I}$, ${\mathbb J}$ and $ {\mathbb D}$ with components
$I_{ijkl} = \frac12 (\delta_{ik}\delta_{jl}+ \delta_{il}\delta_{jk})$, 
$J_{ijkl} = \frac13 \delta_{ij}\delta_{kl}$, and 
$
D_{ijkl} = \delta_{i1}\delta_{j1} \delta_{k1}\delta_{l1} +
\delta_{i2}\delta_{j2} \delta_{k2}\delta_{l2}+
\delta_{i3}\delta_{j3} \delta_{k3}\delta_{l3}$.
A  solid of cubic symmetry has elasticity of the form 
\beq{sc}
{\bf C}  =  3K \, {\mathbb J} + 2\mu_1\, \big( {\mathbb I}- {\mathbb D}\big)
+2\mu_2 \, \big( {\mathbb D}- {\mathbb J}\big) .
\eeq
The isotropic tensor ${\mathbb J}$ and the tensors of cubic symmetry $\big( {\mathbb I}- {\mathbb D}\big)$ and $\big( {\mathbb D}- {\mathbb J}\big)$ are  positive definite \cite{walpole84}, so  the requirement of positive strain energy is that $K$, $\mu_1$ and $\mu_2$ are positive.  These three parameters, called the ``principal elasticities" by Kelvin \cite{kelvin},  can be related to  the  standard Voigt  stiffness  notation: $
K = (C_{11}+2C_{12})/3$, $\mu_1 = C_{44}$ and $\mu_2 = (C_{11}-C_{12})/2$.    The bulk modulus follows from 
eq.\ \eqref{=7-1} as 
\beq{-88}
K= \frac{\phi}9 E \ \ \forall Z; \ \ \
K = \frac{ZR^2}{9VM}, \ Z\ne 14; \ \ \
K = \frac{4a^2}{3V} \big(\frac 1{M_1} + \frac 1{M_2}\big)
, \ Z= 14  
\eeq 
where for $Z=14$ $M_1$, $M_2$ are the axial compliances of 
the two different types of members.   It may be checked that $K_{14}=K_6+K_8$ where $K_Z$ denotes the bulk modulus for coordination number $Z$.   The shear moduli are given in Table \ref{table}. 
Note that the effective compliance, relating strain to stress by $ {\boldsymbol \epsilon} = {\bf C}^{-1}
{\boldsymbol \sigma}$
is simply 
$
{\bf C}^{-1}  =  (3K)^{-1} \, {\mathbb J} + (2\mu_1)^{-1} \, \big( {\mathbb I}- {\mathbb D}\big)
+(2\mu_2)^{-1}  \, \big( {\mathbb D}- {\mathbb J}\big) $.
The ratio $M/N$ may also be expressed in terms of the volume fraction $\phi$ since the rods are assumed to be solid circular    so that 
\beq{003}
\frac MN = \frac 34 \frac{b^2}{R^2}. 
\eeq
Hence, Table \ref{table} indicates that $\mu_1=$O$(\phi^2)$ for $Z=4$, $6$ and 
$\mu_2=$O$(\phi^2)$ for $Z=4$, $8$; otherwise $\mu_1$, $\mu_2=$O$(\phi)$.  

%\FloatBarrier 

\subsubsection*{$Z=14$: the tetrakaidecahedral unit cell}

The tetrakaidecahedron is a truncated octahedron with all edges of the same length  $a$ $\Rightarrow$ $ V=8\sqrt{2}a^3$.   
Rods extend from the center to all faces of the Kelvin cell as shown in Fig.\ \ref{fig1}.  Note the functional dependence $\mu_1 = \mu_1 (M_1, N_2)$, $\mu_2 = \mu_2 (M_2, N_1,N_2)$.  Isotropy $(\mu_1 = \mu_2)$ is achieved 
if 
\beq{85}
\frac 3{M_1} - \frac 3{N_1} = \frac 2{M_2}-\frac 2{N_2} 
\eeq
in which case the effective Poisson's ratio is 
\beq{865}
\nu = \frac{M_1^{-1} - N_1^{-1} } { 4M_1^{-1} - 2N_1^{-1} +2N_2^{-1} }.
\eeq

In the stretch dominated limit $1/N_1 , \, 1/N_2 \to 0$ the $6\times 6$ Voigt matrix of  effective elastic moduli is 
\beq{-26}
{\bf C}_{14} ={\bf C}_6+ {\bf C}_8
\ \ \text{where} \ \
{\bf C}_6 = \frac {\phi_6}{3}E\,  \begin{pmatrix}
{\bf I} & {\bf 0} 
\\
{\bf 0}& {\bf 0}
\end{pmatrix}, 
\ \ 
{\bf C}_8 = \frac {\phi_8}{9} E\, \begin{pmatrix}
{\bf J} & {\bf 0} 
\\
{\bf 0}& {\bf I}
\end{pmatrix} , 
\eeq
all elements of the  $3\times 3$ matrix ${\bf J}$ are unity, 
and   the volume fractions    $\phi_6 = 6R_1A_1/V$, $\phi_8 = 8R_2A_2/V$ satisfy $\phi =\phi_6 + \phi_8 $. 
The  three moduli follow from \eqref{-26} as 
\beq{041}
K =  \frac {\phi}9 E, \ \ 
\mu_1 =  \frac {\phi_8}9 E, \ \ 
\mu_2 =  \frac{\phi_6}6 E .
\eeq
Isotropy, $\mu_1=\mu_2 \equiv \mu$, is achieved if $3\phi_6 = 2 \phi_8$, i.e. 
\beq{3-3}
A_1 = \frac 4{3\sqrt{3}}\, A_2  
\ \  \Rightarrow \ \ 
\mu =  \frac{\phi}{15} E
\eeq
in which case the effective Poisson's ratio is $\nu = \frac 14$, in agreement with \eqref{865}.  This  effective solid  is the 3D isotropic "optimal" material introduced by \cite{Gurtner14}.

\section{Pentamode lattices}  \label{sec5} %%%%%%%%%%%%%%%%%%%%%%%%%%%%%%%%%%%%%%%%%%%%%%%%

\subsection{$Z=d+1$ and the pentamode limit} 

As discussed in \S\ref{sec4}\ref{4.1}, $Z=d+1$ is the minimal coordination number necessary for a fully positive definite elasticity tensor.  We now examine this case in particular in the limit of stretch dominant deformation.  

Given that a PM is an elastic  solid with a single Kelvin modulus the elastic stiffness must be of the form 
\beq{145}
{\bf C} = \lambda \, {\bf S}\otimes {\bf S} , \ \ \lambda >0, \ \     {\bf S}\in\text{Sym}.
\eeq
Note that the parameter $\lambda $ is somewhat arbitrary since it can be replaced by unity by subsuming it into the definition of ${\bf S}$. 
Since rank\,${\bf P}=Z-d$ it follows that the single non-zero eigenvalue of ${\bf P}$ of \eqref{9=2}  is unity, i.e. there exists a $(d+1)$-vector ${\bf b}$ such that 
\beq{9=4}
{\bf P} = {\bf b}{\bf b}^T \ \ \text{where} \ \ {\bf b}^T{\bf b} =1. 
\eeq
Hence, \eqref{9=2}$_1$ yields the moduli explicitly in the form \eqref{145}  with   
\beq{9=5}
%{\bf C} = {\bf S}\otimes   {\bf S} 
%\ \ \text{where} \ \ 
\lambda =1, \ \ 
{\bf S}  = V^{-1/2}\sum_{i=1}^{d+1} R_i M_i^{1/2}\, b_i \, {\bf P}_i^\parallel . 
\eeq
The eigenvalue property  ${\bf P} {\bf b} ={\bf b}$ implies that ${\bf b}$ 
satisfies $ \sum_{i=1}^{d+1} b_i{\bf u}_i = 0$, i.e. it is closely related with the fact that the  $d+1$  vectors ${\bf u}_i$ are necessarily linearly dependent.  
Alternatively, ${\bf b}$ follows by assuming  that ${\bf C}$   of eq.\ \eqref{9=2} has PM form   ${\bf C} = {\bf S}\otimes   {\bf S}$, then use
${\bf C}{\bf I}= {\bf S}\,$tr${\bf S}$ and 
${\bf I}:{\bf C}{\bf I}= ($tr${\bf S})^2$, from which we deduce that the moduli have the form  
\eqref{145} with 
\beq{-11}
\lambda = 
	\big( V
\sum_{k=1}^{d+1} \gamma_k \big)^{-1} ,
\ \ 
{\bf S} = \sum_{i=1}^{d+1}  \gamma_i \,{\bf P}_i^\parallel 
\ \  \text{where}\ \ 
\gamma_i = \frac{R_i^2}{M_i} - \frac {{\bf R}_i }{M_i}  \cdot \Big(
\sum_{k=1}^{d+1} \frac {{\bf P}_k^\parallel}{M_k}   \Big)^{-1}  \cdot 
\sum_{j=1}^{d+1} \frac {{\bf R}_j }{M_j}  .
\eeq
Equations \eqref{145}, \eqref{9=5} and \eqref{-11} provide two alternative and 
explicit formulas for the PM moduli.  

It is interesting to note that either of the above formulas for ${\bf C}$  leads to an  expression for the axial force in member $i$ based on equations  \eqref{2} and \eqref{-39}.   Thus, using eq.\  \eqref{-11} gives 
$ {\bf f}_i = V \lambda ({\bf S}:{\boldsymbol \epsilon}) R_i^{-1}{\gamma_i}  {\bf e}_i$. 
It may  be checked from the definition of $\gamma_i$ that the forces are equilibrated, since 
\beq{-7=}
\sum_{i=1}^{d+1}  R_i^{-1}{\gamma_i}  {\bf e}_i =0 . 
\eeq
This identity implies that $\gamma_i = 0$ for some member $i$ only if (but not iff) the remaining $d$ members are linearly dependent.  When this unusual circumstance occurs the member $i$ bears no load since $ {\bf f}_i = 0$ for any applied strain.  For instance, if two members are collinear in 2D, say members $1$ and  $2$, then 
the third member is not load bearing only if $R_1^{-1}{\gamma_1} =R_2^{-1}{\gamma_2}$.   When $d$ of the members span a $(d-1)$-plane the remaining member is non-load bearing if it is orthogonal to the plane.

%%%%%%%%%%%%%%%%%%%%%%%%%%%%%
 
	   \begin{figure}[h] %%%%%%%%%%%%%%%%%%%%%%%%%%%%%%%%%%%%%%%%%%%%%%%%%%%%%%%%%%
   \begin{center} 
\includegraphics[width=3.in,angle=0]{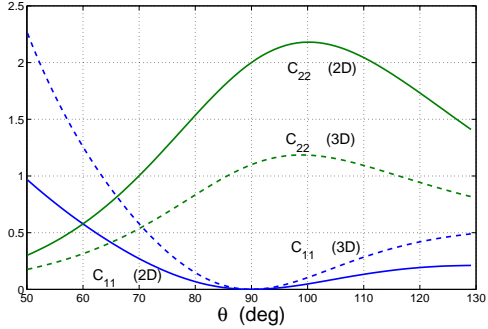}
\caption{The elastic moduli for 2D and 3D PM lattices with rods of equal length $(R_1=R_2)$ and stiffness $(M_1=M_2)$ as a function of the junction angle $\theta$.   Note that the 2D (3D) moduli are identical at the isotropy angle 60$^\circ$ (70.53$^\circ$).  The axial stiffness $C_{11} $ vanishes at $\theta = \frac{\pi}2$.  Since 
$C_{12} = \sqrt{C_{11}C_{22}}$ it follows that $C_{12} $ also vanishes at $\theta = \frac{\pi}2$.   }
   \label{fig3}
	\end{center}
   \end{figure}   %%%%%%%%%%%%%%%%%%%%%%%%%%%%%%%%%%%%%%%%%%%%%%%%%%%%%%%%%%

 Writing  ${\bd S}$ in terms of its  principal directions and eigenvalues, 
$%\beq{046}
 {\bd S} = s_1 {\bd q}_1{\bd q}_1 +s_2 {\bd q}_2{\bd q}_2+s_3 {\bd q}_3{\bd q}_3$,
%\eeq
where $\{ {\bd q}_1, {\bd q}_2,{\bd q}_3\}$ is an orthonormal triad, it follows that 
the  elastic moduli in this basis are 
\beq{3-5}
C_{IJ} = \lambda \, s_Is_J \ \text{if}\  I,J\in\{1,2,3\}, \ 0 \ \text{otherwise}.
\eeq
The material symmetry displayed by  PMs is therefore isotropic, transversely isotropic or orthotropic, the lowest symmetry, depending as the triplet of eigenvalues  
$\{s_1, s_2, s_3\}$ has one, two or three distinct members. 
 The five  ``easy" pentamode strains correspond to the 5-dimensional space 
${\bd S}:{\boldsymbol \epsilon} = 0$.    Three of the easy strains are pure shear: ${\bd q}_i{\bd q}_j +{\bd q}_j{\bd q}_i$, $i\ne j$ and the other two are
$s_1{\bd q}_2{\bd q}_2 -s_2{\bd q}_1{\bd q}_1  $ and 
$s_2{\bd q}_3{\bd q}_3 - s_3{\bd q}_2{\bd q}_2 $.   
Any other zero-energy strain is a linear combination of these.

\subsection{Poisson's ratio of a PM}   

In practice there must be some small but finite rigidity that makes ${\bd C}$ full rank,  the material is unstable otherwise.  The five soft modes of the PM are represented by $0< \{\mu_i,\, i=1, \ldots , 5\} \ll K$ where the set of generalized shear moduli must be determined as part of the full elasticity tensor.   A  measurable quantity that depends upon the  soft moduli is the Poisson's ratio: for a given pair of  directions defined by the orthonormal  vectors 
${\bd n}$ and ${\bd m}$ the Poisson's ratio $\nu_{ nm} $ is  the ratio of the contraction in the ${\bd m}$-direction to the extension in the ${\bd n}$-direction for a uniaxial applied stress along ${\bd n}$, i.e. 
$\nu_{ nm}  = - \big({{\bd m}{\bd m}:{\bd M}{\bd n}{\bd n}} 
\big)/\big({{\bd n}{\bd n}:{\bd M}{\bd n}{\bd n}} \big) $
where ${\bd M} = {\bd C}^{-1}$ is the fourth order tensor of elastic compliance.   
As an example, consider the diamond-like structure of Fig.\  \ref{fig1}(a) with shear  moduli given by Table \ref{table}, $Z=4$. In the pentamode limit $K \gg \mu_1 = 3\mu_2$, with $n_i$, $m_i$ as the components in the principal axes,  
we obtain (see e.g. \cite{Norris05d})
\beq{-02}
\nu_{ nm}  =  \frac{\frac 12 - n_1^2m_1^2- n_2^2m_2^2- n_3^2m_3^2}{n_1^4 +n_2^4 +n_3^4 } 
\in [0,\frac 12] . 
\eeq 

The actual values of the soft moduli $ \{\mu_i,\, i=1, \ldots , 5\}  $ are sensitive to features such as junction strength and might not be easily calculated in comparison with the pentamode stiffness.  An estimate of the Poisson effect can be obtained  by assuming the five soft moduli equal, in which case ${\bd C}(0) \equiv {\bd C}$ of eq.\ \eqref{145} is modified to 
\beq{62}
{\bd C}(\mu)  \equiv
{\bd C}(0) + 2\mu \big( {\mathbb I} - \big(\lambda \text{tr}\, ({\bd S}^2)\big)^{-1}
{\bd C}(0) \big) ,
%\ \  \text{with} \ \tau = \lambda \text{tr}\, ({\bd S}^2) ,
\eeq
which is invertible (and positive definite) for $\mu >0$. 
Using ${\bd M} = {\bd C}^{-1}(\mu)$ define $\nu_{ nm} (\mu)$, then  
the limit exists as the shear modulus is reduced to zero:   $\nu_{ nm} ( 0 ) \equiv \nu_{ nm} $ where 
\beq{66}
\nu_{ nm}  = \frac{ ({\bd m}\cdot{\bd S}{\bd m})({\bd n}\cdot{\bd S}{\bd n}) } 
{ {\bd S}:{\bd S}  -  ({\bd n}\cdot{\bd S}{\bd n})^2}.
\eeq
For the example of Figure  \ref{fig1}(a) ${\bd S}={\bd I}$ and eq.\  \eqref{66} gives $\nu_{ nm} = 1/2$. 
Generally, the values of  $\nu_{ nm} $ from eq.\  \eqref{66}  associated with the principal axes of ${\bd S}$  (see \eqref{3-5}) are
$\nu_{ij} = {s_is_j}/(s_j^2 + s_k^2)$, $i\ne j \ne k \ne i$. 
  If $s_1 > s_2 >s_3 >0$ then the largest and smallest values 
	are $\nu_{12} > \frac 12$ and $\nu_{32} < \frac 12$, respectively. 
Compare this with the Poisson's ratio of an incompressible isotropic elastic material: $\nu = \frac12$.  Negative values of 
Poisson's ratio occur if the principal values of ${\bf S}$ are simultaneously positive and negative.

	   \begin{figure}[H] %%%%%%%%%%%%%%%%%%%%%%%%%%%%%%%%%%%%%%%%%%%%%%%%%%%%%%%%%%
   \begin{center} 
	\subfigure[$\theta=50^\circ$]{%   1.8
\includegraphics[width=1.85in,angle=0]{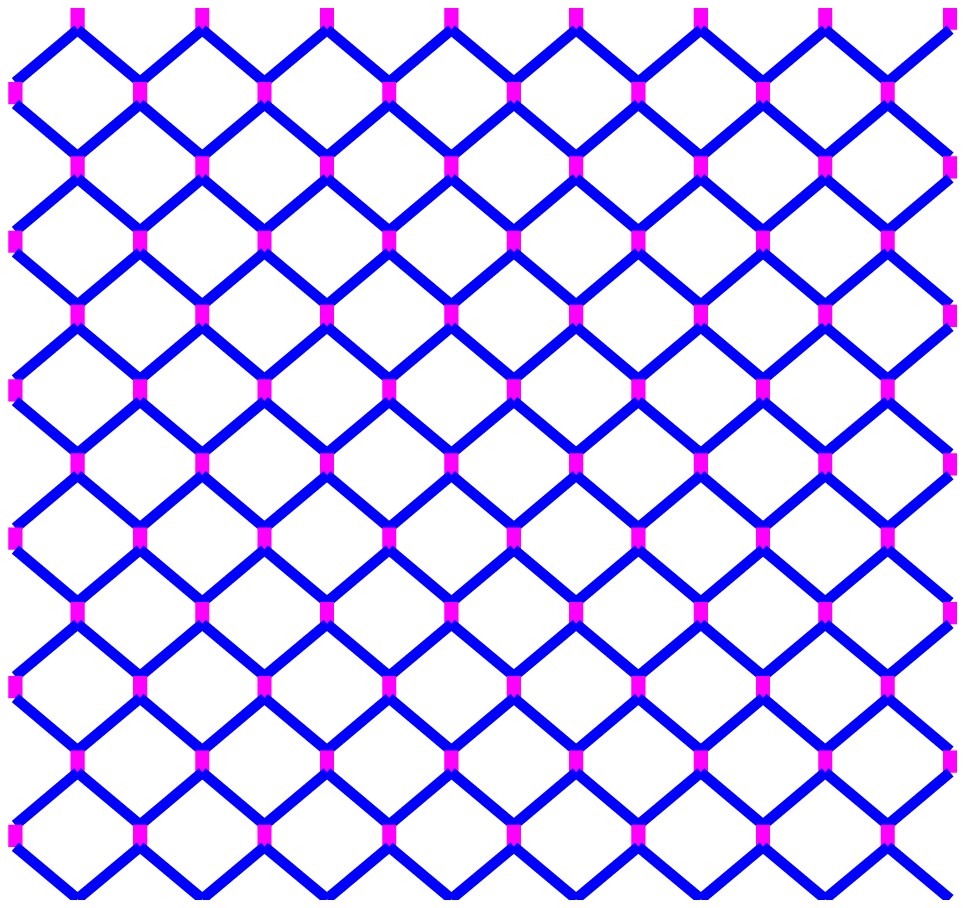}
}
\hspace{-0.35in}
	\subfigure[$\theta =60^\circ$]{%
\includegraphics[width=1.85in,angle=0]{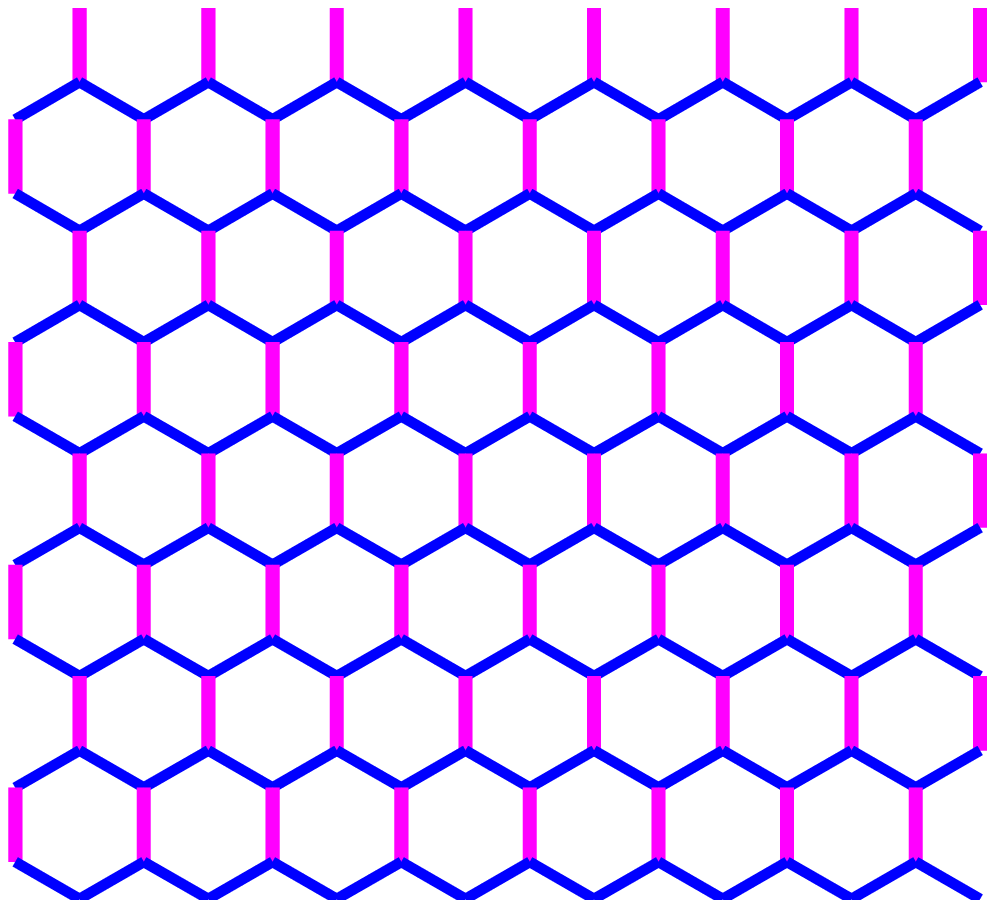}
}\hspace{-0.32in}
	\subfigure[$\theta =70^\circ$]{%
\includegraphics[width=1.85in,angle=0]{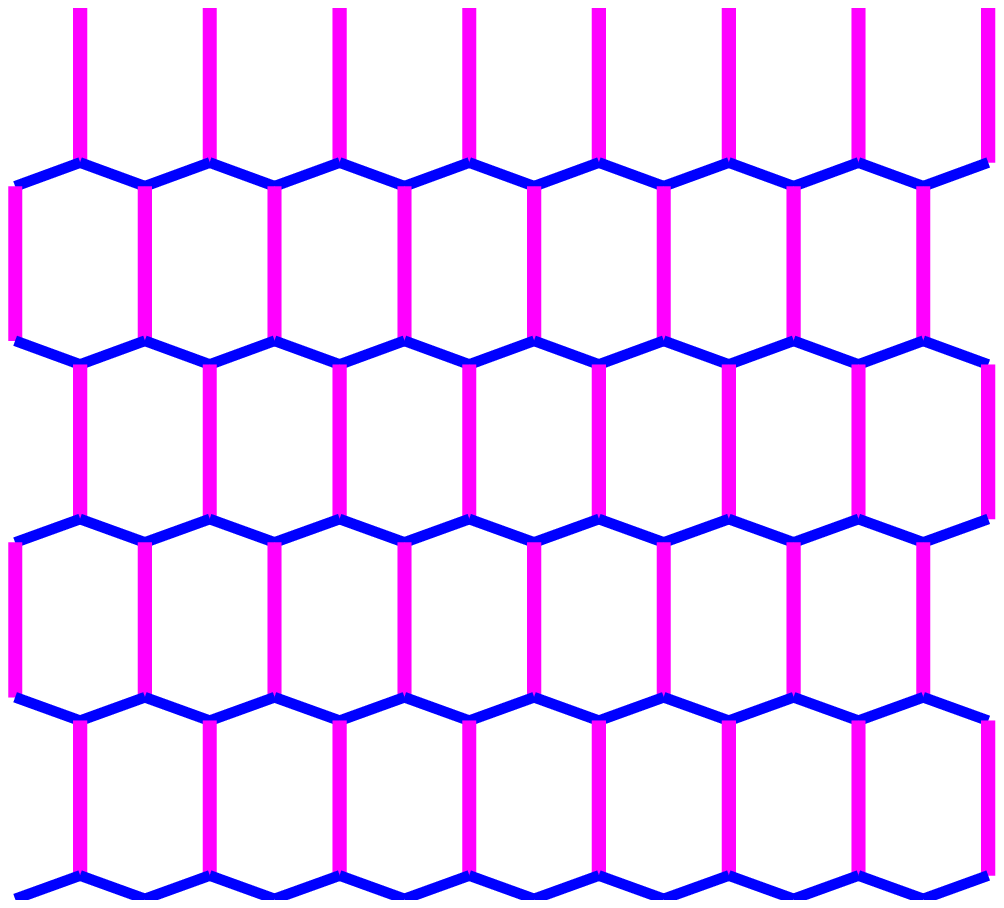}
}
\caption{Each of these two-dimensional  PM lattices have isotropic quasi-static properties.  The ratio of the $R_1$ (red)  to $R_2$ (blue) is determined by \eqref{4-1}. The pure honeycomb structure is  $\theta =60^\circ$. }
   \label{fig2}
	\end{center}
   \end{figure}   %%%%%%%%%%%%%%%%%%%%%%%%%%%%%%%%%%%%%%%%%%%%%%%%%%%%%%%%%%

		   \begin{figure}[h] %%%%%%%%%%%%%%%%%%%%%%%%%%%%%%%%%%%%%%%%%%%%%%%%%%%%%%%%%%
   \begin{center} 
		\subfigure[$\nu_{12}, \, \nu_{21}$]{
\includegraphics[width=3.in,angle=0]{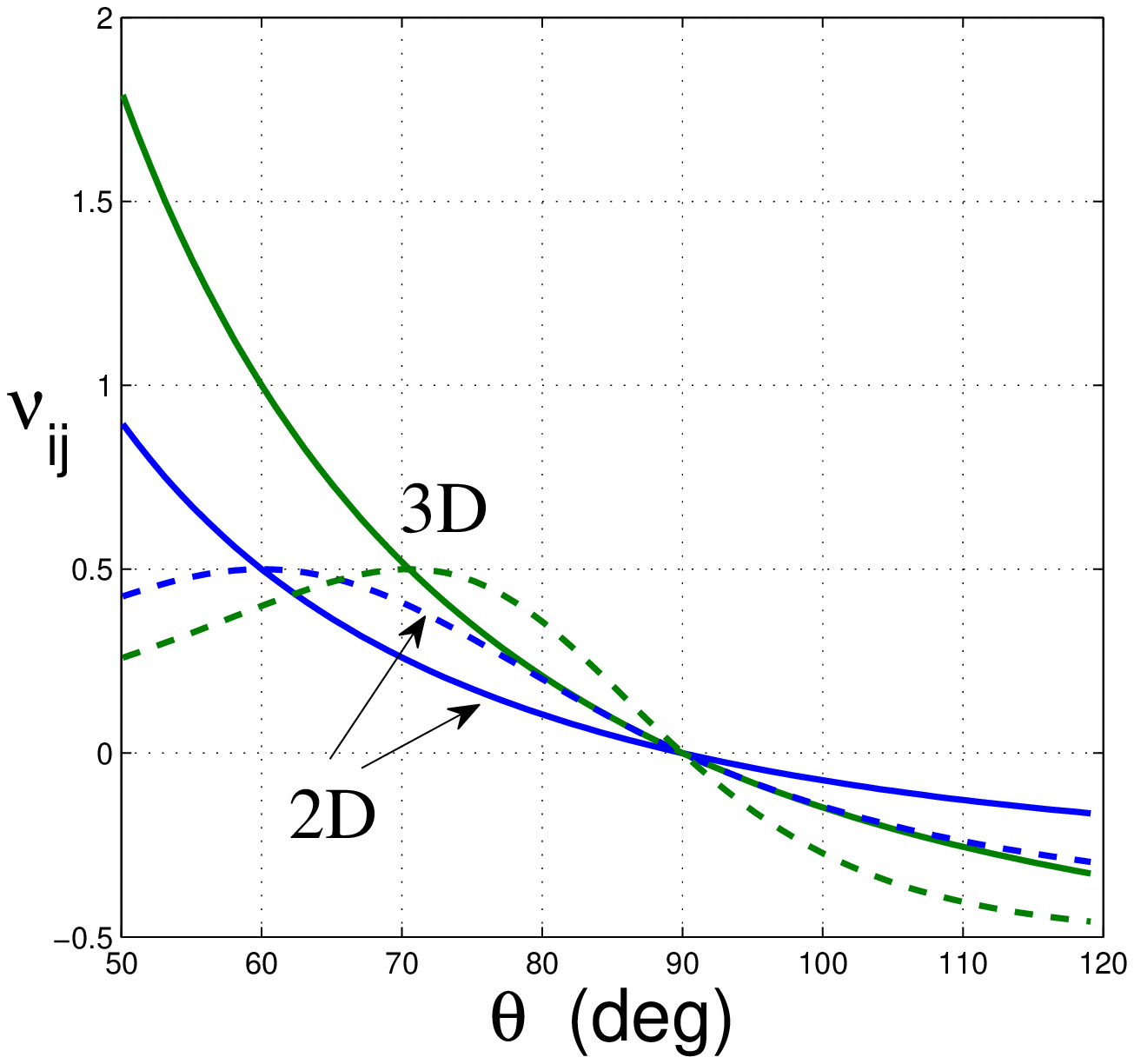}
}
\hspace{-0.35in}
	\subfigure[2D: $\theta =50^\circ , \, 110^\circ$]{
 \begin{tabular}[b]{c}
\includegraphics[width=1.8in,angle=0]{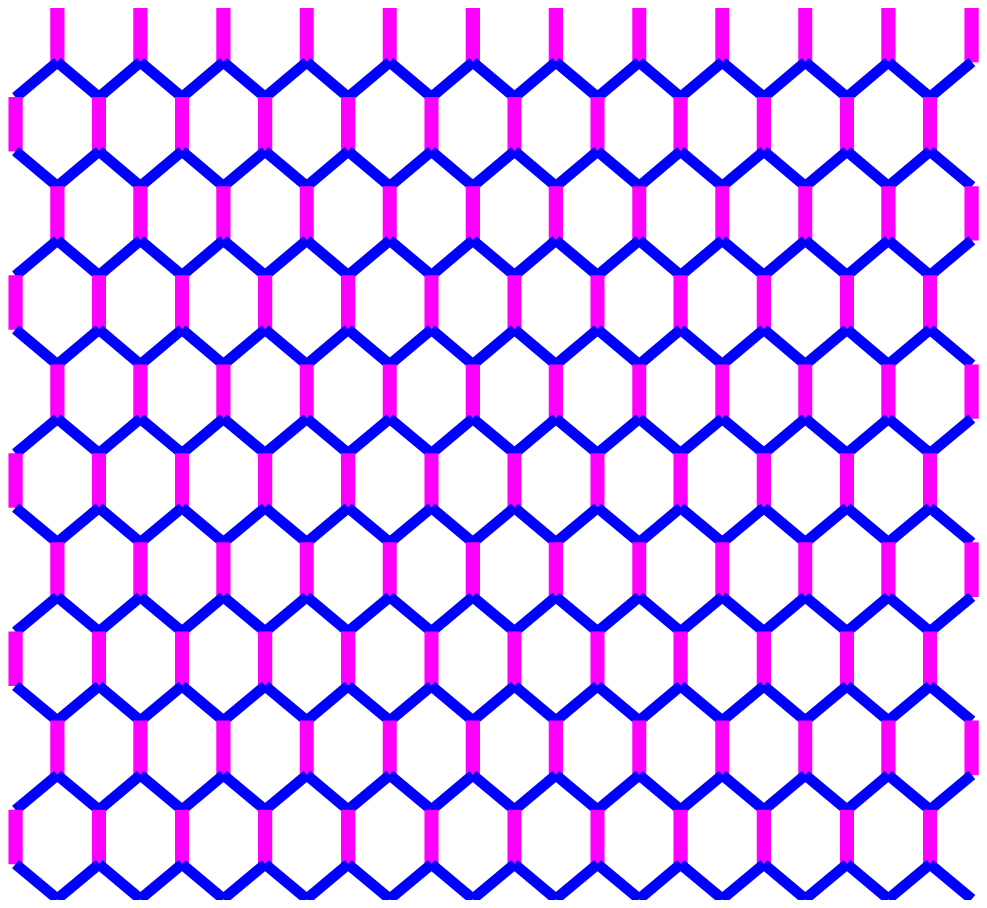}
\\ %[2mm] 
\includegraphics[width=1.5in,angle=0]{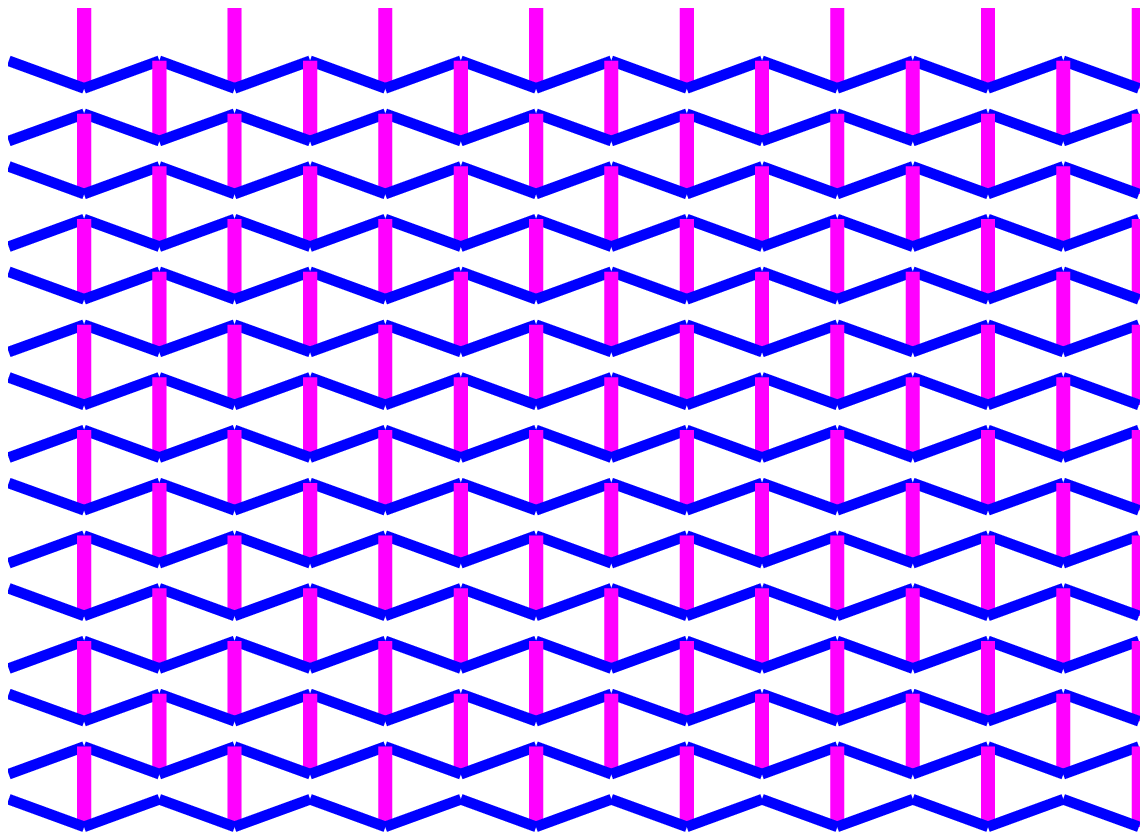}
   \end{tabular}
}	
\caption{(a) The solid curves show Poisson's ratio $\nu_{12}$ for the same configuration as Figure \ref{fig3} ($R_1=R_2$, $M_1=M_2$).  $\nu_{12}$ %This Poisson's ratio 
describes the lateral contraction for  loading along the axial ${\bf e}$-direction.   The related Poisson's ratio 
 $\nu_{21} = \nu_{12}/( \frac 12 + 2 \nu_{12}^2)$ is shown by the dashed curves. 
(b) The 2D lattice for $\theta =50^\circ$ (top) and $\theta= 110^\circ$.    }
   \label{fig4}
	\end{center}
   \end{figure}   %%%%%%%%%%%%%%%%%%%%%%%%%%%%%%%%%%%%%%%%%%%%%%%%%%%%%%%%%%
	
				   \begin{figure}[h] %%%%%%%%%%%%%%%%%%%%%%%%%%%%%%%%%%%%%%%%%%%%%%%%%%%%%%%%%%
   \begin{center} 
		\subfigure{ %[$C_{11},\, C_{22}$]{
\includegraphics[width=2.5in,angle=0]{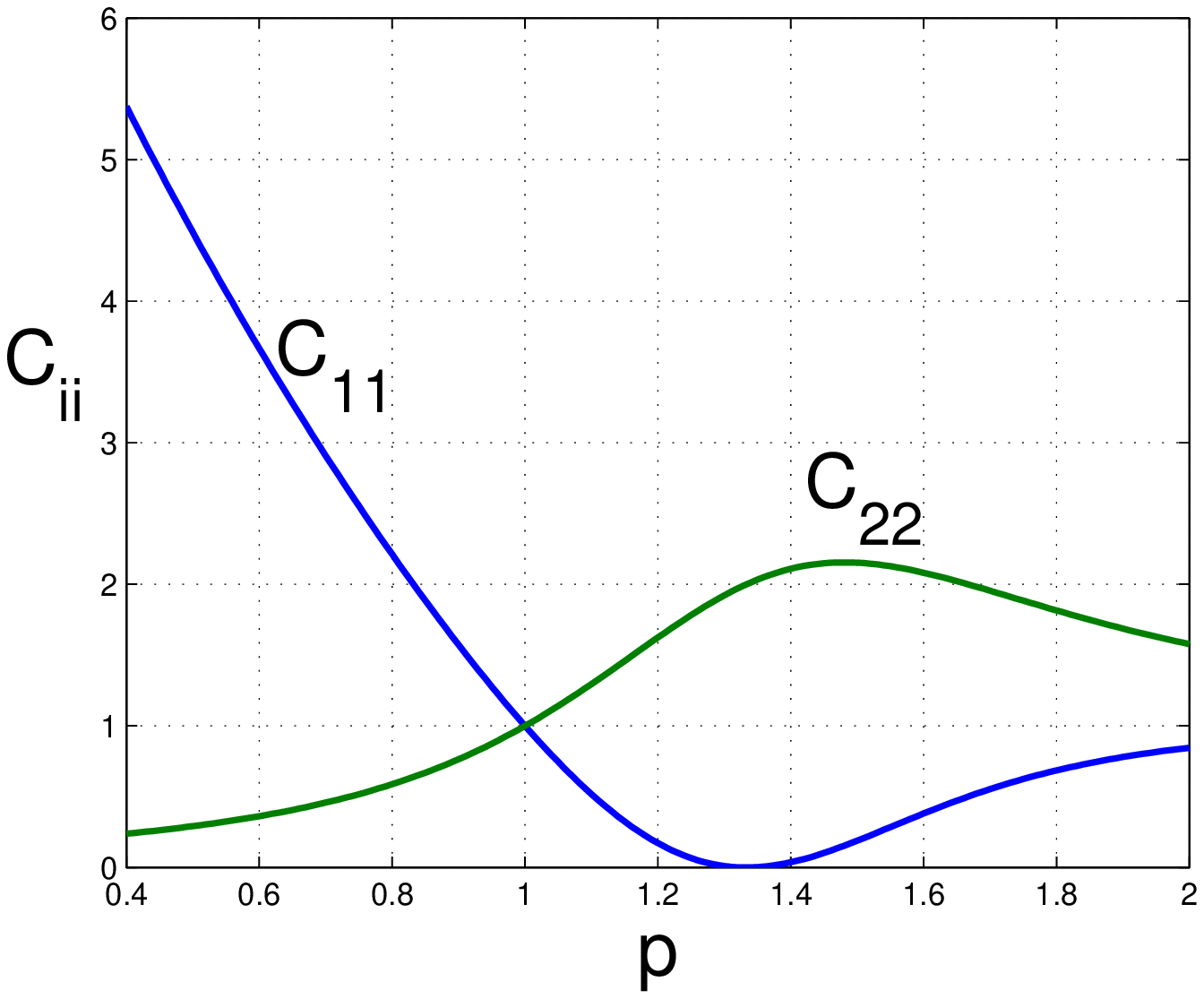}
}
\hspace{-0.2in}
	\subfigure{% [$\nu_{12}, \, \nu_{21}$]{
 \includegraphics[width=2.5in,angle=0]{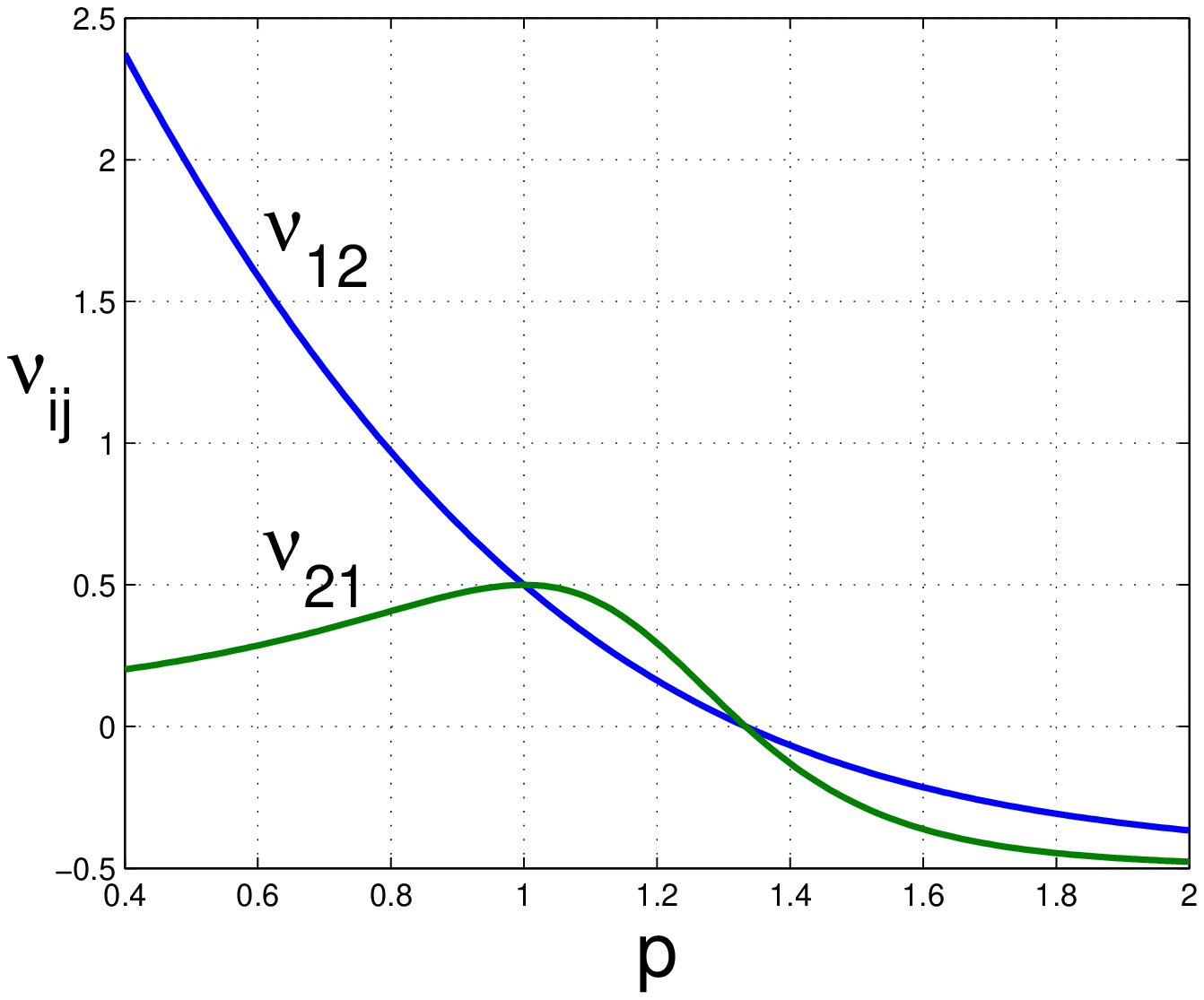}
}	
\caption{The principal stiffnesses (left) and Poisson's ratios (right) for a diamond lattice with the center "atom" shifted along the cube diagonal.   The four vertices of the unit cell at $(0\, 0\, 0)$, $(0\, 2\, 2)$, $(2\, 0\, 2)$,  $(2\, 2\, 0)$, and the center junction (atom) lies at $(p\, p\, p)$.  Isotropy is $p=1$.}
   \label{fig5}
	\end{center}
   \end{figure}   %%%%%%%%%%%%%%%%%%%%%%%%%%%%%%%%%%%%%%%%%%%%%%%%%%%%%%%%%%

\subsection{Transversely isotropic PM lattice} \label{1.3.2}

Assume the unit cell has symmetry consistent with macroscopic transverse isotropy.  It comprises two types of rods: $i=1$ with  $R_1$, $M_1$ in direction ${\bf e}$ $(={\bf e}_1)$, and $i=2,\ldots , d+1$ with  $R_2$, $M_2$ in directions ${\bf e}_i$  symmetrically situated about $-{\bf e}$ with  $-{\bf e}\cdot {\bf e}_i = \cos \theta$.   Let $c=\cos \theta$, $s=\sin\theta$.  
We find after some simplification, that \eqref{145} and \eqref{-11} give the PM elastic stiffness as 
\beq{17}
{\bf C} =\frac {ds^4 R_2^2} {V (d-1)^2  (dc^2 M_1 +M_2)}
\big(  {\bf I}  +   (\beta -1){\bf e}\otimes {\bf e}  \big)
\otimes 
\big(  {\bf I}  +   (\beta -1){\bf e}\otimes {\bf e}  \big)
\eeq
where the non-dimensional parameter $\beta$ and the unit cell volume $V$ are  
\beq{44}
\beta = \frac {(d-1)c \big(R_1+cR_2\big) } { s^2 R_2 }, 
\quad 
V =   (sR_2 )^{d-1} \big(R_1 +    c R_2 \big) \,\times \begin{cases}
4,& d=2,
\\
6\sqrt{3}, & d=3.
\end{cases} 
\eeq
Note that the elasticity of the rods enters only through the combination $dc^2 M_1 +M_2$.  

The nondimensional  geometrical parameter $\beta$ defines the anisotropy of the pentamode material, with isotropy iff $\beta = 1$.  If $\beta >1$ the  PM is stiffer along the axial or preferred direction ${\bf e}$ than in the orthogonal plane, and conversely it is stiffer in the plane if $0< \beta <1$.  The axial stiffness vanishes if $\beta = 0$ which is possible if $\theta = \frac{\pi}2$.  The unit cell becomes re-entrant if $\theta > \frac{\pi}2 \Leftrightarrow c<0$.  If $c<0$  then $\beta <0$ and  the principal values of ${\bf S}$ are simultaneously positive and negative with the negative value associated with the axial direction.  Note that  $R_1+cR_2$ must be positive since the unit cell volume $V$ is positive.  As $R_1+cR_2 \to 0$  the members criss-cross  and the infinite lattice becomes stacked in a slab of unit thickness, hence the volume per cell tends to zero  $(V\to 0)$.

Let ${\bf e}$, the axis of transverse isotropy,  be in the $1$-direction.  A transversely isotropic elastic solid $(d=3)$ has 5 independent moduli: $C_{11}$, $C_{22}$ $(=C_{33})$, $C_{12}$ $(=C_{13})$, $C_{23}$  and 
$C_{66}$ $(=C_{55})$ with $C_{44} = \frac 12 (C_{22}-C_{23})$.  The PM  has $C_{66}=0$ and $C_{23}=C_{22}$ $(\Rightarrow C_{44}=0)$ and $C_{11} C_{22} = C_{12}^2$, which are consistent with rank\,${\bf C}=1$.  The 2D version, technically of orthotropic symmetry, is defined by 4 independent moduli $C_{11}$, $C_{22}$, $C_{12}$   and $C_{66}$, which in the PM limit satisfy $C_{66}=0$ and $C_{11} C_{22} = C_{12}^2$.  In either case the non-zero moduli are   
\beq{19}
\begin{pmatrix}
C_{11} & C_{12} 
\\
C_{12}  & C_{22} 
\end{pmatrix}
= K_0
\begin{pmatrix}
\beta  &1 
\\
1  & \beta^{-1} 
\end{pmatrix}
\ \ \text{where} \ 
K_0 =  \frac {d}{(d-1)}
 \frac  {c s^2 R_2 \big(R_1+cR_2\big) }{V (M_2+ {dc^2} M_1)}  .
\eeq
%%%%%%%%

The PM is isotropic for  $\beta = 1$, i.e. when the angle $\theta$ and $R_1/R_2$ are related by 
\beq{4-1}
\frac{R_1}{R_2} = \frac{1- d\cos\theta^2}{(d-1)\cos\theta}
\ \ \Leftrightarrow \ \ \text{isotropy} \ \ (\beta = 1). 
\eeq
Hence, isotropy can be obtained if 
$ \theta \in [\cos^{-1}\frac1{\sqrt{d}},\, \frac{\pi}2]$ 
with the proper ratio of lengths, see Fig.\ \ref{fig2}.  At  the limiting angles 
$R_1 \to 0$ $(R_2 \to 0)$ as $ \theta \to \cos^{-1}\frac1{\sqrt{d}}$ 
$ (\theta \to \frac{\pi}2 )$.  If the lengths are equal $({R_1}={R_2} )$ isotropy is obtained  for $\cos\theta  = \frac 1d $, i.e. 
  $\theta = $60$^\circ$, 70.53$^\circ$, for $d=2$, $3$,  corresponding to hexagonal and tetrahedral unit cells, respectively.    Some examples of isotropic PMs and their properties are illustrated in Fig.\ \ref{fig2}.   Transversely isotropic PMs are considered in Figs.\  \ref{fig3} - \ref{fig5}.

\begin{figure}[h] %%%%%%%%%%%%%%%%%%%%%%%%%%%%%%%%%%%%%%%%%%%%%%%%%%%%%%%%%%
   \begin{center} 
	\includegraphics[width=1.5in,angle=0]{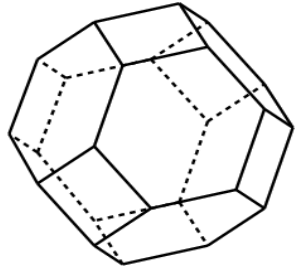}
\caption{The tetrakaidecahedral open foam unit cell \cite{Warren97} has low density PM behavior similar to the diamond lattice. }
   \label{fig14sides}
	\end{center}
   \end{figure}

The stiffness parameter $K_0$ of \eqref{19} is the bulk modulus of the isotropic PM.  Note that $K_0$ is not equivalent to $K$ of \eqref{=7-1} since the latter is consequent upon the condition \eqref{999} which is not assumed here.  Instead, eqs.\ \eqref{=12},  \eqref{19}   imply that the isotropic PM bulk modulus for  uniform members is 
\beq{04=}
K_0=  K f, %\big(\theta, d, \frac{A_1}{A_2}\big), 
\ \   f = d^2s^4\big[ d-1 + \frac{A_1}{A_2dc}(1-dc^2)\big]^{-1}
             \big[ d-1 + \frac{A_2}{A_1}dc(1-dc^2)\big]^{-1}
\eeq
where $A_1$, $A_2$ are the cross-sectional areas (strut thicknesses for $d=2$).  For a given $\theta$ and $d$, $f\le 1$ with equality iff $\frac{A_1}{A_2} = dc$.   Hence  the maximum possible isotropic effective bulk modulus for a given volume fraction $\phi$ is precisely $K$ of \eqref{=7-1}.  
This result agrees with \cite[eq.\ (2.2)]{Christensen95} for $d=2$, and with the bulk modulus for a regular lattice with tetrakaidecahedral unit cells  \cite{Warren97},  \cite{Zhu97}, i.e. an open Kelvin foam, see Fig.\  \ref{fig14sides}.  The latter structure,  comprising joints with 4 struts and a unit cell of 14 faces (6 squares and  8 hexagons), has cubic symmetry; however the two shear moduli are almost equal so that the structure is almost isotropic.  In fact, if the struts are circular and have Poisson's ratio equal to zero then the effective material is precisely isotropic with shear modulus $\mu =  \frac{4\sqrt{2}}{9\pi}\phi^2 E$ \cite{Warren97}.

	      %%%%	
Note that \cite{Warren88} considered a tetrahedral unit cell of four identical half-struts that join at equal angles and found $K= \frac{\phi E}{8}$ (not $\frac{\phi E}{9}$); the difference  arises from taking the cell volume  for the tetrahedron, but since the tetrahedron is not a space filling polyhedron, this is not the correct unit volume to use.

\section{Two dimensions: a special case}  \label{sec6}

\subsection{Shear force as a nodal bending force}  

For $d=2$ the total  force \eqref{-45} on  member $i$ can be simplified as 
\beq{5=}
{\bf f}_i  = M_i^{-1} \Delta r_i  {\bf e}_i 
+ \sum_{j\ne i} {N_{ij}'}^{-1}  R_j  \, \Delta \psi_{ij}\,  {\bf e}_{ij} %{\boldsymbol \theta}_{ij}
\eeq
with   generalized nodal compliance $N_{ij}'$  given by 
\beq{5=1}
\frac1{N_{ij}'} = \frac1{N_{ij}} + \frac1{N^{(b)}_{ij}} 
\ \  \text{where} \ 
N^{(b)}_{ij} \equiv  \frac{N_iN_j} {R_iR_j} \,   \sum_k \frac{R_k^2}{N_k}   
.
\eeq
Hence, the shear force can be considered as an equivalent nodal  bending force.  Significantly, the moments of the shear forces are now automatically equilibrated due to the symmetry  $N_{ij}' = N_{ji}'$. 

Equation \eqref{5=} follows by first noting that the vector moment of the shear force is in the direction perpendicular to the plane of the lattice, say ${\bf a}_3$.  Define the angle of deflection associated with flexural bending: $\theta_i \equiv  {\bf a}_3 \cdot ({\bf e}_i \times \Delta  {\bf r}_i^b )/R_i $.  The moment of the shear force is    ${\bf R}_i \times  {\bf f}_i^b = (R_i^2/N_i)\theta_i {\bf a}_3$, and the 
moment equilibrium condition \eqref{4=7} becomes
\beq{=8}
\sum_i \frac{R_i^2}{N_i}  \theta_i =0 
\ \ \Rightarrow \ \ 
\theta_i = \Big(\sum_k \frac{R_k^2}{N_k} \Big)^{-1} \sum_{j\ne i} \frac{R_j^2}{N_j} (\theta_i -\theta_j).
\eeq
However, $\theta_i -\theta_j = \pm \Delta\psi_{ij}$ (more precisely    $\theta_i -\theta_j = \Delta\psi_{ij}
\,  {\bf a}_3 \cdot ({\bf e}_j \times {\bf e}_i)/|{\bf e}_j \times {\bf e}_i| $), therefore eq.\ \eqref{=8} allows us to express the single shear force acting on member $i$ as the sum of nodal bending forces with compliances $N^{(b)}_{ij}$, from which eq.\ \eqref{5=} follows.  

The significance of eq.\ \eqref{5=} is that it allows us to express the effective moduli for $d=2$ as follows: 
Define
\bse{11-}
\bal{7=11}
{\bf d}_i &= \frac{ {\bf e}_i}{\sqrt{M_i}} ,  
 \ \ 
{\bf D}_i  =  R_i 
\frac{{\bf e}_i \otimes {\bf e}_i}{\sqrt{VM_i }}  ,  
\ \ 
{\bf d}_{ij} = \sqrt{ \frac{R_iR_j}{N_{ij}'}} \,  
\Big( \frac{{\bf e}_{ij}}{R_i} +\frac{{\bf e}_{ji}}{R_j} \Big) , 
\\
{\bf D}_{ij}   &= 
 \sqrt{ \frac{R_iR_j}{VN_{ij}'}} \,  
\big( 
{\bf e}_i\otimes  {\bf e}_{ij}
+ {\bf e}_j\otimes {\bf e}_{ji} \big) 
\ \text{where} \ 
\frac1{N_{ij}'} = \frac1{N_{ij}} +  \frac{R_iR_j}{N_iN_j}  \,  
\Big( \sum_{k=1}^Z \frac{R_k^2}{N_k}  \Big)^{-1} , 
\\
\left. \{ {\bf u}_k \}\right|_{k=1}^L &= \{  {\bf d}_i , \, {\bf d}_{ij}  \}, 
\ \ 
\left. \{ {\bf U}_k \}\right|_{k=1}^L = \{  {\bf D}_i , \, {\bf D}_{ij}  \} , 
\ \ L = Z(Z+1)/2, 
\\
P_{ij} &= \delta_{ij} - {\bf u}_i\cdot \Big(
\sum_{k=1}^{N}  {\bf u}_k\otimes {\bf u}_k  \Big)^{-1} \cdot {\bf u}_j
\quad \Rightarrow \quad 
{\bf C} =   \sum_{i, j=1}^{L}
 P_{ij}  {\bf U}_i\otimes {\bf U}_j  .
\eal
\ese
Note that this result is valid for any similarly situated 2D lattice structure; in particular it does not require the zero rotation assumption \eqref{3=0}.    

\subsection{Example:  honeycomb lattice}

As an application of eq.\ \eqref{11-} we consider the transversely isotropic lattice  of \S\ref{sec5}\ref{1.3.2} in 2-dimensions $(Z=3)$, now including the effects of the bending compliances of the individual  members,  $N_1$ and $N_2$.    Using the same notation as in \S\ref{sec5}\ref{1.3.2} we find
\bal{002}
\ba
\left. \begin{matrix}
C_{11} 
\\
C_{22} 
\\
C_{12} 
\end{matrix}\right\} &=
\frac {\frac 12 cs  }{ 
 (2c^2 M_1+M_2)N_2 + 2 s^2 M_1 M_2}
	\times
	\begin{cases}
	\beta \big( N_2 + s^2 c^{-2} M_2 \big) ,
	\\
\frac 1{\beta}  \big( N_2 + s^{-2} (2M_1 + c^2M_2) \big) ,
 \\
  (N_2- M_2)  ,
	\end{cases}
\\
C_{66} &=
 \frac { \frac 12 s R_2  
 \left( R_1+c R_2 \right)    }
{    s^2 \big(2R_2^2N_1+ R_1^2N_2\big) + \big( cR_1+ R_2 \big) ^2 M_2 }
.
\ea
\eal
 These  are in agreement with the in-plane moduli found by \cite{Kim03}.  Note that the moduli of eq.\ \eqref{002}  reduce to the PM moduli \eqref{19} as the bending compliance $N_2 \to \infty$, independent of the bending compliance $N_1$.

\section{Conclusions}  \label{sec7}

Our main result, eq.\ \eqref{11}, is that the  effective moduli of the lattice structure can be expressed 
$ {\bf C} =   \sum_{i, j=1}^{L}
 P_{ij}  {\bf U}_i\otimes {\bf U}_j 
$ where  $L =Zd + Z(Z-1)/2$,  $ {\bf U}_i$ are  second order tensors, and $P_{ij} $ are elements of a $L\times L$ projection matrix ${\bf P} $ %$\big(={\bf P}^2\big) $
of rank $L-d$.  
Explicit forms for the parameters  $\{ {\bf U}_i,\, P_{ij}\}$
have been  derived in terms of the cell volume, and the  length,   orientation, axial and bending stiffness of each of the  $Z$ rods.  This Kelvin-like representation for the elasticity tensor 
 implies as a necessary although not sufficient condition for positive definiteness of ${\bf C}$  that the rank of ${\bf P}$ exceed $3d-3$, which  is satisfied  if the coordination number satisfies $Z \ge d+1 $.  The $L$ second order tensors $\{ {\bf U}_i  \}$ are split into 
$Z$ stretch dominated and $Z(Z-1)/2$ bending dominated elements.  The latter contribute little to the stiffness in the limit of very thin members, in which case the elastic stiffness is stretch dominated and, at most,  of rank $Z-d$.  The formulation developed here is applicable to the entire range of stiffness possible in similarly situated lattice frameworks, from the $Z=14$ structure proposed by \cite{Gurtner14} with full rank   ${\bf C}$ to  pentamode materials corresponding to coordination number  $Z=d+1$, with  ${\bf C}$ of  rank one.

\section*{Acknowledgment}
%\ack
{Thanks to Adam Nagy and Xiaoshi Su for discussions and graphical assistance. This work  was supported under ONR MURI Grant No. N000141310631}

%%%%%%%%%%%%%%%%%%%%%%%%%%%%%%%%%%%%%%%%%%%%%%%%%%%%%%%%%%%%%%%%%%%%%%%%%%
%\bibliography{../../../SHARED_BIBLIOGRAPHY/AN_BIG_BIB}
%\bibliographystyle{bibgen}%ieeetr}%abbrv}%plain}%unsrt}%unsrtnat}%natbib}%unsrtnat}%doipubmed}%harvard}% plain}%uabbrvnat}%
%
%\end{document}

\end{document}